\documentclass[a4paper,11pt]{article}
\pdfoutput=1 

\usepackage{jheppub} 

\usepackage[T1]{fontenc} 

\usepackage{braket}
\usepackage{todonotes}
\usepackage{braket}
\usepackage{appendix}
\usepackage{float}
\usepackage{array}
\usepackage{comment}
\usepackage{xcolor}
\usepackage{amsmath}
\title{\boldmath Spectral Form Factor of Gapped Random Matrix Systems}

\author[1]{Krishan Saraswat}

\affiliation[1]{Physics Department, Broida Hall, University of California, Santa Barbara, CA 93106, USA}

\emailAdd{ksaraswat@ucsb.edu}

\abstract{In this work, we study the spectral form factor of random matrix models which exhibit a large number of degenerate ground states accompanied by a macroscopic gap in the spectrum. The central aim of this work is to understand how the standard narrative about the behavior of the spectral form factor is modified in the presence of these parametrically large number of ground states. We show that, at sufficiently low temperatures, the spectral form factor is dominated by the disconnected contribution, even at arbitrarily late times. Moreover, we demonstrate that the connected form factor only depends on the eigenvalues of the non-degenerate sector, implying that BPS states do not contribute to wormhole calculations in the gravity context. Using the Christoffel–Darboux kernel, we analyze a number of examples including the Bessel model and $\mathcal{N}=2$ Jackiw–Teitelboim supergravity. In these examples, we find damped oscillations in the disconnected form factor, with a period set by the inverse size of the gap. Furthermore, we demonstrate that the slope of the ramp in the connected form factor arises from a universal sine-kernel, which emerges from a truncation of the full non-perturbative kernel in the $\hbar\to 0$ limit, and find agreement with the leading double trumpet result. Finally, we present predictions for how the ramp will transition to a plateau in the connected form factor and demonstrate how the transition depends on the details of the leading spectral density of states.}

\begin{document} 
\maketitle

\section{Introduction}
The study of the spectral form factor has its origins in the field of random matrix theory \cite{wishart1928generalised,hsu1939distribution,wigner1951statistical,wigner1955characteristic,wigner1958distribution,dyson1962threefold,marchenko1967distribution,tracy1994airy,tracy1996orthogonal,mehta2004random,forrester2010log,livan2018introduction}. One of the earliest uses of the spectral form factor was to study certain universal aspects of eigenvalue correlations and chaos in random matrix ensembles \cite{berry1985semiclassical,bohigas1984characterization,berry1977level,haake2010quantum,PhysRevE.55.4067,PhysRevLett.78.2280}. More recently, over the past few decades, it has been shown that black holes and various aspects of quantum gravity (especially 2D quantum gravity) can be understood in terms of random matrix models \cite{Gross:1989aw,Ginsparg:1993is,Altland_1997,Shenker:2013pqa,Maldacena:2015waa,Cotler:2016fpe,Magan:2018nmu,Sarosi:2017ykf,Almheiri:2020cfm,Johnson:2019eik,Johnson:2020heh,Johnson:2022wsr,Turiaci:2024cad}, and with it, a renewed interest in the spectral form factor in the context of quantum gravitational and black hole physics \cite{Cotler:2016fpe,Saad:2019lba,Saad:2019pqd,Johnson:2020exp,Johnson:2020mwi,Johnson:2021owr,Saraswat:2021ong,Belin:2021ibv,Choi:2022asl,Saad:2022kfe,Blommaert:2022lbh,Bhattacharyya:2023gvg,Okuyama:2023pio,Anegawa:2023klh,Ageev:2024gem}. One of the earliest works which made use of the spectral form factor to study quantum chaotic aspects of black hole microstates was \cite{Cotler:2016fpe}, which computed the spectral form factor for the Sachdev-Ye-Kitaev (SYK) model\footnote{This model was relevant in the study of black hole microstates since it was shown to be related to 2D Jackiw–Teitelboim gravity (a gravity theory describing near horizon physics of certain near-extremal black holes) in the IR regime \cite{Cotler:2016fpe,Sarosi:2017ykf}.}. There, the spectral form factor could be expressed in terms of energy eigenvalues of the SYK Hamiltonian and written explicitly as an expression of the form: 
\begin{equation}
    Z(\beta+it)Z(\beta-it)=\sum_{n,m}e^{-\beta(E_n+E_m)}e^{it(E_m-E_n)}\ .
\end{equation}
Which is a 2-point function of the operator $e^{-\beta H}$. Furthermore, it has also served as an interesting proxy to study the 2-point functions of more general simple operators, $O$, in the thermal ensemble which can be expressed as:
\begin{equation}
\begin{split}
    &\text{Tr}\left[e^{-\beta H}O(t)O(0)\right]=\text{Tr}[e^{-\beta H}e^{-iHt}O(0)e^{iHt}O(0)]\\
    &=\sum_{i,j}\braket{i|e^{-\beta H}e^{-iHt}O\ket{j}\bra{j}e^{iHt}O|i}=\sum_{i,j}e^{-\beta E_i}e^{i(E_i-E_j)t}|\braket{i|O|j}|^2\ .\\
\end{split}
\end{equation}
As one can see, the time dependence of thermal correlators of $O$ come in the form of a phase which depends on the difference between microstate energies weighted by matrix elements between microstates. Due to this fact, many features of the the late time behavior of the form factor can analogously appear in correlators of simple operators as well, as was demonstrated in \cite{Cotler:2016fpe}. In the context of black hole physics, such correlators would contain the physics of how perturbations thermalize when a black hole is perturbed \cite{Horowitz:1999jd,Maldacena:2001kr,Herzog:2002pc,Festuccia:2005pi}. For this reason, there have been a number of subsequent works which regard the spectral form factor as a simpler proxy than the 2-point function of simple operators to study black hole thermalization \cite{Balasubramanian:2016ids,Collier:2021rsn,Saraswat:2021ong}.

In recent years, there has been a great deal of interest and progress towards understanding black hole microstates in supergravity especially in the context of Jackiw–Teitelboim (JT) supergravity \cite{Iliesiu:2020qvm,Heydeman:2020hhw,Boruch:2022tno,Iliesiu:2022kny,Boruch:2023trc,Cassani:2025sim}. A particularly remarkable feature of the low energy spectrum of microstates in such theories is the existence of a large number\footnote{Here, by ``large'' we mean that the number of BPS states scales with $ e^{S_0}$, where $S_0$ is the black hole entropy. Also note, that throughout this paper we often use $\hbar$ rather than $e^{-S_0}$ these two describe the same expansion parameter in the genus expansion of the gravity theory.} of degenerate Bogomol'nyi–Prasad–Sommerfield (BPS) states at zero energy as well as a macroscopic gap which forms as a result of enhanced microstate repulsion between the non-BPS and BPS states \cite{Johnson:2024tgg}. Generally, the appearance of such degenerate microstates and gaps is not generic, but rather appears due to supersymmetry in these special class of theories. Perhaps due to this, most of the examples that have studied the spectral form factor in the literature (to our knowledge) have focused on random matrix models with no macroscopic gaps present. The purpose of this work is to fill in this ``gap'' (pardon the pun). In particular, we will investigate how the presence of these large number of degenerate states and the gap appear in computations of the spectral form factor. Although our goal will be to eventually gain insights about the form factor for gravitational theories with a gapped spectrum our results are expected to apply more generally to any system whose spectrum has the properties discussed above.  

In \textbf{Section \ref{GenAspFFSection}}, we begin by reviewing the standard narrative and generic aspects about the spectral form factor (including its decomposition into connected and disconnected parts) under the assumption that no substantial degeneracies occur in the spectrum. This is then contrasted to the case which involves a spectrum with a large number of degenerate zero energy states. A central insight we gain through this new analysis is that, if the number of degenerate states is comparable to the number of non-degenerate states, the disconnected part of the spectral form factor will not decay to zero and always remains as the dominant contribution when compared to the connected form factor (even at arbitrarily late times). In addition, we also prove that the connected part of the form factor only takes input from the set of non-degenerate states in the random sector. Nonetheless, we argue that although there is no explicit dependence, there are still subtle ways by which the degenerate sector makes itself known in the connected form factor which we illustrate through the examples considered in this work.

In \textbf{Section \ref{SFFWishartSection}}, we analyze the spectral form factor of matrices belonging to the Wishart ensemble. We begin by giving a brief review of the Christoffel–Darboux kernel (or simply kernel) and its relation to computations to the spectral form factor. Through our subsequent analysis of the disconnected form factor we uncover the existence of damped oscillations after the initial dip of the form factor whose asymptotic period converges towards a time scale exactly given by $\tau=\frac{2\pi}{E_{Gap}}$, where $E_{Gap}$ is the size of the gap in the spectrum. Using a variety of numerical and analytic techniques we also study the subleading connected form factor and demonstrate how the slope of the ramp of the connected form factor is sensitive to the presence of the degenerate sector and provide estimates on how the onset time of the plateau depends on the number of degenerate eigenvalues.

In \textbf{Section \ref{BesselModFFSection}}, we consider an example of a double scaled matrix model called the ``Bessel Model'', which appears in the scaling limit of the Wishart ensemble near the left edge of the spectrum of states. We begin this section by giving a quick review of a general formalism which allows to one construct the kernel of a large class of double scaled matrix models (including the Bessel model). Using the kernel, we study both the connected and disconnected form factor. As in the Wishart example, the disconnected form factor also contains oscillations whose period is exactly characterized by the gap size in the spectrum. However, only at sufficiently ``low'' temperatures (i.e. when $\beta\gg \left(\frac{\hbar}{E_{Gap}}\right)^2$) will the disconnected part dominate at all time scales. In our analysis of the connected form factor we make use of the full non-perturbative kernel of the Bessel model and demonstrate the emergence of a universal sine-kernel (given in Eq. \ref{BesselToSineKernel}) which depends on the leading density of states as $\hbar\to 0$. Employing this universal form, we explicitly study the ramp and plateau of the connected form factor and provide closed form expressions for the onset time of the plateau and even provide a prediction for the slope of the ramp, both of which depend on the number of degenerate states. A particularly satisfying outcome of the analysis is the agreement of the slope of the ramp using the Kernel approach with slope predicted by gluing two trumpets at leading order in the genus expansion.

In \textbf{Section \ref{N=2Section}}, we discuss the example of $\mathcal{N}=2$ JT supergravity making use of the same formalism as we did for the Bessel model\footnote{In recent work \cite{Johnson:2026plw}, it was demonstrated that the Bessel model also lives within $\mathcal{N}=2$ JT supergravity near the left edge of the spectrum as well.}. Just as in previous examples, the disconnected part at sufficiently low temperatures will be dominant and exhibit oscillations whose period is governed by the gap between the BPS and non-BPS part of the spectrum. In addition, we conjecture that, one should expect a universal sine-kernel to emerge from the full non-perturbative description of the kernel of $\mathcal{N}=2$ JT supergravity in much the same way as was demonstrated explicitly in the Bessel model. Using this conjecture, we demonstrate how to reproduce the leading double trumpet (sometimes called the ``wormhole'') result and also demonstrate that, unlike the Bessel model, there is not a sharp transition from the ramp to plateau phase. We also do a numerical analysis of the form factor using the sine-kernel truncation and compare the result to the leading double trumpet result and show very close agreement during the ramp phase (e.g. see Figure \ref{N=2ConnFFConject}).

We conclude in \textbf{Section \ref{ConcludeSec}} by discussing key results of the work and provide interesting new avenues of exploration.

\section{General Aspects of the Spectral Form Factor}
\label{GenAspFFSection}
\subsection{Form Factor of Random Matrix Systems}
In preparation to study the spectral form factor of systems with large degeneracies and spectral gaps, it will be useful to review well known aspects of the spectral form factor in the absence of degenerate sub-sectors (i.e. the standard story of the form factor). Any reader who is already familiar with the standard narrative of how the form factor behaves at early and late times is invited to skip this subsection and go directly to the next subsection which discusses how things get modified in the presence of degenerate sub-sectors. 

The spectral form factor is defined as the following product of analytically continued partition functions \cite{Cotler:2016fpe,Saad:2019lba,Saad:2019pqd,Johnson:2020exp,Johnson:2020mwi,Johnson:2021owr,Saraswat:2021ong,Belin:2021ibv,Choi:2022asl,Saad:2022kfe,Blommaert:2022lbh,Okuyama:2023pio,Anegawa:2023klh,Ageev:2024gem}:
\begin{equation}
    Z(\beta+it)Z(\beta-it)\ ,
\end{equation}
where $Z(\beta)$ is the canonical partition function of the system at inverse temperature $\beta$. For a system living in a finite dimensional Hilbert space we can write $Z(\beta)=\sum_{n=1}^N e^{-\beta E_n}$ and explicitly write:
\begin{equation}
    Z(\beta+it)Z(\beta-it)=\sum_{n,m=1}^Ne^{-\beta(E_n+E_m)}e^{it(E_m-E_n)}\ .
\end{equation}
In the case of random matrices or ensemble averages we are interested in computing the the average over many samples and define the ensemble averaged form factor as:
\begin{equation}
    \braket{Z(\beta+it)Z(\beta-it)}=\lim_{N_{\text{Trial}} \to \infty}\frac{1}{N_{\text{Trial}}}\sum_{I=1}^{N_{\text{Trial}}}\left[\sum_{n,m=1}^Ne^{-\beta(E^{(I)}_n+E^{(I)}_m)}e^{it(E^{(I)}_m-E^{(I)}_n)}\right]\ ,
\end{equation}
 where $E_n^{(I)}$ is the $n$-th eigenvalue of the $I$-th trial. We decompose the expression for the spectral form factor into diagonal and off-diagonal parts as follows:
 \begin{equation}
 \begin{split}
     &\braket{Z(\beta+it)Z(\beta-it)}\\
     &=\lim_{N_{\text{Trial}} \to \infty}\frac{1}{N_{\text{Trial}}}\sum_{I=1}^{N_{\text{Trial}}}\left[\sum_{n=1}^Ne^{-2\beta E_n^{(I)}}+\sum_{\substack{n, m=1 \\ n \neq m}}^Ne^{-\beta(E^{(I)}_n+E^{(I)}_m)}e^{it(E^{(I)}_m-E^{(I)}_n)}\right]\\
     &=\braket{Z(2\beta)}+\lim_{N_{\text{Trial}} \to \infty}\frac{1}{N_{\text{Trial}}}\sum_{I=1}^{N_{\text{Trial}}}\sum_{n>m}\sum_{m=1}^N 2e^{-\beta(E^{(I)}_n+E^{(I)}_m)}\cos\left[(E_n^{(I)}-E_m^{(I)})t\right]\ .\\
 \end{split}
 \end{equation}
 Based on the expression above we can make a few comments about the behavior of the form factor at $t=0$ and very late times, $t\to \infty$. At very late times the off diagonal terms will wildly oscillate with random phase about an average value given by $\braket{Z(2\beta)}$. As the number of trials goes to infinity the amplitude of the oscillations will decay to zero yielding a smooth plateau of height $\braket{Z(2\beta)}$. In particular, at $\beta=0$ the height of the plateau will be~$\lim_{\beta\to 0}\braket{Z(2\beta)}=N$, which is the total number of states of the system. On the other hand, at time $t=0$, the value of the spectral form factor will be $\braket{Z(\beta)^2}$ and will always satisfy $\braket{Z(2\beta)} \leq \braket{Z(\beta)^2} \leq N^2$ the lower bound can be seen by noting that the off diagonal parts of the form factor at $t=0$ are always positive and the upper bound is obtained by noting that the exponential function (over positive reals) is maximized when~$\beta=0$\footnote{Here we made an implicit assumption that the lowest energy state has a non-negative energy. In fact we can make it general and simply require the system has a lowest energy state(s).}.

In many examples involving random matrix ensembles (e.g. GOE, GUE, and GSE) the form factor typically includes three major phases \cite{Cotler:2016fpe,Liu:2018hlr}. The first phase called the ``dip'' describes an initial decay of the form factor and is ``self-averaging'' (i.e. the form factor of a single draw from the ensemble is well approximated by the average). In particular, in the dip phase once can approximate the full ensemble averaged form factor with a related quantity called the disconnected averaged form factor (or simply disconnected form factor) which is defined by:
\begin{equation}
\begin{split}
    & \braket{Z(\beta+it)Z(\beta-it)}_{\text{dis.}}=\braket{Z(\beta+it)}\braket{Z(\beta-it)} \\
    &=\lim_{N_{\text{Trial}}\to\infty}\left[\frac{1}{N_{\text{Trial}}}\sum_{I=1}^{N_{\text{Trial}}}\sum_{n=1}^Ne^{-(\beta+it)E_n^{(I)}}\right]\left[\frac{1}{N_{\text{Trial}}}\sum_{J=1}^{N_{\text{Trial}}}\sum_{m=1}^Ne^{-(\beta-it)E_m^{(J)}}\right]\ , \\
\end{split}
\end{equation}
the major difference being that instead of taking the product of partition functions and then doing an average we instead take averages over each copy of the partition function. We can see that at $t=0$ the disconnected form factor is equal to $\braket{Z(\beta)}^2$. For the late time analysis we ignore the set of non-diagonal terms that oscillate with time and focus on the $I=J$ and $n=m$ terms to obtain a value of $N^{-1}_{\text{Trial}}\braket{Z(2\beta)}$ which will go to zero as~$N_{\text{Trial}}\to \infty$. This is expected since as the number of samples goes to infinity we should converge towards some continuous density of states $\rho(E)$ (satisfying $\int \rho(E)dE=N<\infty$) and we can effectively write:
\begin{equation}
    \braket{Z(\beta+it)}\braket{Z(\beta-it)}=\left(\int dE \rho(E)e^{-(\beta+it)E}\right)\left(\int dE \rho(E)e^{-(\beta-it)E}\right)\ , 
\end{equation}
where the integrals will vanish as $t\to \infty$ by the Riemann–Lebesgue lemma.

The next phase after the dip is called the ``ramp'' in which the value of the form factor increases linearly in $t$. The origins of this ramp lie in the repulsion statistics between pairs of eigenvalues in the spectrum which is a typical signature of quantum chaos. The final phase of the form factor is the plateau in which the form factor is constant. Unlike the initial dip, the ramp and plateau phases of the form factor are not self averaging and a single draw from the ensemble will fluctuate wildly around the ensemble averaged result. Both the ramp and plateau phase of the ensemble averaged form factor can be understood in terms of a quantity called the ``connected'' contribution to the form factor (or just the connected form factor) which is simply given by the difference between the full ensemble averaged form factor and the its disconnected part:
\begin{equation}
    \braket{Z(\beta+it)Z(\beta-it)}_{\text{con}.}=\braket{Z(\beta+it)Z(\beta-it)}-\braket{Z(\beta+it)}\braket{Z(\beta-it)}\ .
\end{equation}
By these definitions, the full form factor is a sum of connected and disconnected contributions. Based on this, we can see that as the number of trials goes to infinity the late time behavior is dominated by the connected part of the form factor due to the fact that the disconnected part goes to zero and since the full ensemble averaged form factor will attain a value equal to $\braket{Z(2\beta)}$ it implies that $\lim_{t\to \infty}\braket{Z(\beta+it)Z(\beta-it)}_{con.}=\braket{Z(2\beta)}$. On the other hand, when $t=0$ we can see that the connected part of the spectral form factor will take on a value given by $\braket{Z(\beta)^2}-\braket{Z(\beta)}^2\geq 0$ which is the variance of the partition function, generally it will be zero at infinite temperature and non-negative at finite temperature. Below we provide a table summarizing the values of the average form factor(s) at times $t=0$ and $t\to \infty$.

\begin{table}[ht!]
    \centering
   \begin{tabular}{ l|c|c } 
  & $t=0$ & $t\to\infty$ \\ 
 \hline
 $\braket{Z(\beta+it)Z(\beta-it)}$ & $\braket{Z(\beta)^2}$ & $\braket{Z(2\beta)}$ \\ 
 $\braket{Z(\beta+it)Z(\beta-it)}_{\text{dis.}}$ & $\braket{Z(\beta)}^2$ & 0 \\ 
 $\braket{Z(\beta+it)Z(\beta-it)}_{\text{con.}}$ & $\braket{Z(\beta)^2}-\braket{Z(\beta)}^2$ & $\braket{Z(2\beta)}$ \\ 
\end{tabular}
    \caption{This table shows the values of the various averaged form factors we defined thus far and gives their values of $t=0$ and very late time ($t\to\infty$).} 
    \label{FFValuesNoDegens}
\end{table}
A particularly important assumption made in these computations is that all the eigenvalues in the spectrum are random variables which exhibit eigenvalue repulsion. However, we are interested in the case where a subset of eigenvalues are non-random (i.e. the degenerate sector). The goal of the next subsection is to do an analogous analysis as we reviewed here but with the assumption that there are additional non-random sectors of eigenvalues present. 

\subsection{Spectral Form Factor in the Presence Degenerate Sub-sectors}
\label{SFFGenericAnalysisRandom+NonRandSec}
The results given in this subsection will generally be true for any system whose spectrum is a mixture of non-random and random sectors where the random sector arises from a random matrix ensemble. Suppose we have a square random matrix of size $N+\Gamma$ whose $N$ eigenvalues vary randomly as we consider various draws from the ensemble. The remaining~$\Gamma$ set of eigenvalues are not random, in the sense that from every single draw of the ensemble the eigenvalues do not vary\footnote{A concrete example of a ensemble that does this one that is generated by a rectangular random matrix~$M$. We can construct a square random matrix by considering either $MM^\dagger$ or $M^\dagger M$ in one of these combinations there will be a degenerate number of eigenvalues which will not be random. Although in this particular example all non-random eigenvalues are degenerate the results of this section will apply even if the non-random eigenvalues are not degenerate.}. Let us denote the $n$-th eigenvalue from the $I$-th draw of the ensemble as $E_n^{(I)}$. Without loss of generality we will take the first $N$ eigenvalues of a particular draw from the ensemble to be the random eigenvalues with the remaining~$\Gamma$ ones being non-random. With this, the ensemble averaged spectral form factor is:
\begin{equation}
\begin{split}
    &\braket{Z(\beta+it)Z(\beta-it)}=\lim_{N_{\text{Trial}}\to \infty}\frac{1}{N_{\text{Trial}}}\sum_{I=1}^{N_{\text{Trial}}}\left[\left(\sum_{n=1}^{\Gamma+N}e^{-(\beta+it)E^{(I)}_n}\right)\left(\sum_{m=1}^{\Gamma+N}e^{-(\beta-it)E^{(I)}_m}\right)\right]\\
    &=\lim_{N_{\text{Trial}}\to \infty}\frac{1}{N_{\text{Trial}}}\sum_{I=1}^{N_{\text{Trial}}}\left[\left(Z_{NR}(\beta+it)+Z_R(\beta+it)\right)\left(Z_{NR}(\beta-it)+Z_R(\beta-it)\right)\right]\\
    &=\braket{Z_{NR}(\beta+it)Z_{NR}(\beta-it)}+\braket{Z_{NR}(\beta+it)Z_R(\beta-it)}+\braket{Z_{NR}(\beta-it)Z_{R}(\beta+it)}\\
    &+\braket{Z_R(\beta+it)Z_R(\beta-it)}\\
    &Z_{NR}(\beta)=\sum_{n=N+1}^{N+\Gamma} e^{-\beta E_n}, \qquad Z_R(\beta)=\sum_{n=1}^{N} e^{-\beta E^{(I)}_n}\ ,\\
\end{split}
\end{equation}
where we simply expanded the form factor into partition functions associated with the random eigenvalues $Z_{R}(\beta)$ and the non-random eigenvalues $Z_{NR}(\beta)$. Note that in the definition of $Z_{NR}(\beta)$ we have $E_n^{(I)}=E_n$ for $n=N+1,N+2,..,N+\Gamma$ and all $I\in\mathbb{N}$. Due to this, the ensemble average acts trivially over any factors of $Z_{NR}$. This leaves us with the following expression for the form factor in the presence of non-random sub-sectors:
\begin{equation}
\begin{split}
    &\braket{Z(\beta+it)Z(\beta-it)}=Z_{NR}(\beta+it)Z_{NR}(\beta-it)+Z_{NR}(\beta+it)\braket{Z_R(\beta-it)}\\
    &+Z_{NR}(\beta-it)\braket{Z_{R}(\beta+it)}+\braket{Z_R(\beta+it)Z_R(\beta-it)}\ . \\
\end{split}
\end{equation}
Using similar methods we can also compute the averaged disconnected part of the form factor as follows:
\begin{equation}
    \begin{split}
        &\braket{Z(\beta+it)Z(\beta-it)}_{\text{dis.}}=\braket{Z(\beta+it)}\braket{Z(\beta-it)}\\
        &=\braket{Z_{NR}(\beta+it)+Z_R(\beta+it)}\braket{Z_{NR}(\beta-it)+Z_R(\beta-it)}\\
        &=\left(Z_{NR}(\beta+it)+\braket{Z_R(\beta+it)}\right)\left(Z_{NR}(\beta-it)+\braket{Z_R(\beta-it)}\right)\\
        &=Z_{NR}(\beta+it)Z_{NR}(\beta-it)+Z_{NR}(\beta+it)\braket{Z_R(\beta-it)}+Z_{NR}(\beta-it)\braket{Z_{R}(\beta+it)}\\
        &+\braket{Z_R(\beta+it)}\braket{Z_R(\beta-it)}\ .\\
    \end{split}
\end{equation}
From these results the connected form factor is given as:
\begin{equation}
\begin{split}
    &\braket{Z(\beta+it)Z(\beta-it)}_{\text{con.}}=\braket{Z_R(\beta+it)Z_R(\beta-it)}_{\text{con.}}\\
    &=\braket{Z_R(\beta+it)Z_R(\beta-it)}-\braket{Z_R(\beta+it)}\braket{Z_R(\beta-it)}\ .\\
\end{split}
\end{equation}
We see that terms explicitly involving the non-random sectors completely cancel. This is a key result that demonstrates that the connected part of the form factor is solely determined by the random sector and will obey similar properties of the connected form factor reviewed in the previous subsection. In the context of the gravitational path integral for systems where there are non-random BPS states (e.g. $\mathcal{N}=2$ JT supergravity), this implies that the famous ``wormhole/double trumpet'' configuration which connects two boundaries is insensitive to the presence of the BPS states - it only ``knows'' about the random non-BPS sector of microstates. This is not just true for 2-boundary wormholes but also for any $N>2$ boundary wormhole configurations (which is completely connected), demonstrated explicitly in Appendix \ref{NboudaryWHProof}.

Now consider the special case of interest to us, namely that the non-random sector has exactly $\Gamma$ number of degenerate states with energy $E=0$. In this case, we have~$Z_{NR}(\beta)=\Gamma$. Substituting this into our expressions allows us to construct Table \ref{FFValuesWithDegens}, which lists the values for spectral form factor(s) at time $t=0$ and $t\to\infty$.
\begin{table}[ht!]
    \centering
   \begin{tabular}{ l|c|c } 
  & $t=0$ & $t\to\infty$ \\ 
 \hline
 $\braket{Z(\beta+it)Z(\beta-it)}$ & $\Gamma^2+2\Gamma \braket{Z_R(\beta)}+\braket{Z_R(\beta)^2}$ & $\Gamma^2+\braket{Z_R(2\beta)}$ \\ 
 $\braket{Z(\beta+it)Z(\beta-it)}_{\text{dis.}}$ & $\left(\Gamma+\braket{Z_R(\beta)}\right)^2$ & $\Gamma^2$ \\ 
 $\braket{Z(\beta+it)Z(\beta-it)}_{\text{con.}}$ & $\braket{Z_R(\beta)^2}-\braket{Z_R(\beta)}^2$ & $\braket{Z_R(2\beta)}$ \\ 
\end{tabular}
    \caption{This table shows the values of the various averaged form factors $t=0$ and very late time ($t\to\infty$) assuming there are $\Gamma$ number of non-random degenerate states at $E=0$.} 
    \label{FFValuesWithDegens}
\end{table}

One of the most striking results of the entries of Table \ref{FFValuesWithDegens} is the last column describing the late time behavior of the spectral form factor. In particular, the value of the disconnected part of the form factor will not decay to zero and this value can be made to be arbitrarily large compared to the height of the plateau of the connected part of the form factor. This result suggests that, in regimes where $\Gamma^2\gg \braket{Z_R(2\beta)}$, the connected part of the spectral form factor will never dominate (even at late times), it always remains as a sub-leading contribution\footnote{This observation has potentially interesting consequences for the factorization puzzle \cite{Maldacena:2004rf,Saad:2019lba,Blommaert:2021fob,Iliesiu:2021are}. Effectively, the result suggests that the gapped random matrix systems we are interested in (which includes $\mathcal{N}=2$ JT supergravity) approximately factorize in the sense that the non-factorizing connected part is subdominant compared to the factorizing disconnected part on all time scales. More specifically, this will happen in $\mathcal{N}=2$ JT supergravity at low enough temperatures.}. This in stark contrast to the standard lore summarized in Table \ref{FFValuesNoDegens} where the disconnected part decays to zero and becomes sub-dominant at late times. One may ask if regimes where $\Gamma^2\gg \braket{Z_R(2\beta)}$ are physically interesting to our discussion. The answer to this is a resounding yes! In fact, for the examples we consider in this work, this regime is not just a mathematical curiosity it is forced upon us when discussing random matrix systems with macroscopic gaps. This can be understood from the work \cite{Johnson:2023ofr,Johnson:2024tgg}, where it was shown that to obtain a macroscopic gap one needs $\Gamma=\tilde{\Gamma}N$, where $N$ describes the number of non-degenerate excited states and $\tilde{\Gamma}$ is some order $1$ parameter. Based on this, it is clear that since $\braket{Z(2\beta)}\sim N$ and $\Gamma^2\sim N^2$ we are firmly in the regime where~$\Gamma^2\gg \braket{Z_R(2\beta)}$ at sufficiently large $N$. The same regime will also appear at sufficiently low temperatures as well\footnote{This statement is particularly important in double scaled matrix models where $N\to\infty$. In this case, the plateau at a finite temperature has a height which measures an effective number of excited states, $N_{\text{eff}}<\infty$. This can be lowered as the temperature is lowered, and in particular, lowered enough so that $N_{\text{eff}}\ll \Gamma^2$.}. So the result that the disconnected part is in fact the dominant part of the calculation is not something we can simply ignore in such systems. Furthermore,  The next important observation to make from our analysis is that the connected part of the spectral form factor only explicitly depends on the randomly distributed eigenvalues and not on the eigenvalues in the degenerate sectors. What this implies is that when one computes the connected form factor in such settings it is only giving information about the microstate statistics of the randomly distributed sector. 

In light of these observations, we can see from Table \ref{FFValuesWithDegens}, the explicit $\Gamma$ dependence comes purely from the self averaging disconnected part. At first glance, this appears rather uninteresting, in the sense that usually the most interesting part of the form factor is the connected part, and naively the inclusion of the degenerate sector does not affect the connected part. This would be a valid point if we view the non-random eigenvalues as simply being passively put in by hand with no affect on the random sector. However, this is not true. We should properly view the random and non-random sector as emerging from diagonalizing a larger random matrix. Since the two sectors emerge from the same underlying random matrix we should expect changes in one sector to affect the other. A prime example of this in action is the spectral gap that appears due to enhanced eigenvalue repulsion \cite{Johnson:2024tgg}. As we will see, this not the only way the non-degenerate sector can be affected, the spectral density of the non-degenerate sector will also deform as we change the number of states in the degenerate sector. As we will demonstrate in this work, this will generally manifest in the connected form factor by changing the slope of the ramp and onset time of the plateau phase.  

Furthermore, the the question of whether the connected versus the disconnected part of the form factor is of interest may also depend on what kind of physics or aspects of the system we are trying to study. If we are purely interested in aspects of the eigenvalue statistics themselves then the connected form factor is more important regardless of whether it is dominant or not. However, if we wish to view the spectral form factor as a simple proxy to study the thermalization physics of a system \cite{Balasubramanian:2016ids,Collier:2021rsn,Saraswat:2021ong}, then the connected form factor only represents the thermalization physics associated to transitions between states in the random sector and can potentially miss interesting physics describing possible transitions between degenerate and non-degenerate sectors. In fact, one may suggest that in the regimes we are interested in, the dominant physics describing thermalization is coming from studying the disconnected part rather than the sub-leading connected part. 

Regardless of how one wants to view the form factor, we hope to convey that in random matrix systems with macroscopic gaps both the connected and disconnected parts of the spectral form factor are potentially interesting quantities to compute with each of them containing some interesting features pertaining to the physics of the gap and the degenerate sub-sector. In what follows, we will employ many of the examples used in the work \cite{Johnson:2024tgg} and study the spectral form factor to gain insight into the details of how the degenerate states make themselves known.

\section{Spectral Form Factor of Wishart Ensembles}
\label{SFFWishartSection}
In this section, we will be discussing the spectral form factor of the Wishart ensemble. In the past, there have been some works that have studied the form factor of matrices belonging the Wishart ensemble \cite{Liu:2018hlr,Hunter-Jones:2017crg,Forrester_2021}. These studies tend to focus on the version of the ensemble where there are no degenerate zero eigenvalues. To our knowledge, we are not aware of works that explicitly discuss features of the spectral form factor of Wishart ensembles which includes a large number of degenerate zero eigenvalues which we will study in this section \footnote{In particular, the work \cite{Forrester_2021}, does consider a parameter regime which we are interested in (where the bulk of the density is shifted away from the origin). However, it does not explicitly include the zero eigenvalues. Due to the results of Section \ref{SFFGenericAnalysisRandom+NonRandSec} much of what is discussed in \cite{Forrester_2021} in regards to the connected form factor remains unchanged, but there will be significant divergence in the discussions of the disconnected form factor.}.     
\subsection{Matrices and Kernel of the Wishart Ensemble}
In this subsection, we give a brief review of the matrices involved in the Wishart ensemble as well as the standard Christoffel–Darboux kernel and its relation to the spectral form factor. We will then use these results in subsequent subsections to study the disconnected and connected form factors. 

Matrices in the Wishart ensemble can be constructed by introducing a rectangular $N\times (N+\Gamma)$ matrix, $Q$ \cite{wishart1928generalised,hsu1939distribution,marchenko1967distribution,livan2018introduction}. For concreteness, we will consider the matrix $Q$ to have complex entries of the form $x+iy$ and assuming $x,y$ to be real parameters which are Gaussian distributed\footnote{Similar to the classical matrix ensemble GOE, GUE, and GSE \cite{dyson1962threefold} a similar categorization can also be made with the Wishart ensemble in terms of the real, complex, and symplectic Wishart ensembles \cite{forrester1993laguerre}.}. We can then use such rectangular random matrices to construct two classes of Hermitian matrix. The first choice is $W_-=QQ^\dagger$ which results in a $N\times N$ Hermitian random matrix whose non-negative eigenvalues will all be randomly distributed. The other choice corresponds to $W_{+}=Q^\dagger Q$ which will be a $(N+\Gamma)\times(N+\Gamma)$  Hermitian random matrix. The positive eigenvalues of $W_+$ and $W_{-}$ are identical but $W_+$ has $\Gamma+N-N=\Gamma$ number of degenerate zero eigenvalues\footnote{Although in this paper we simply refer to both matrices $W_+$ and $W_-$ as belonging to the Wishart ensemble there are instances in literature where an explicit distinction is made between the two \cite{Yu_2002,Janik:2001cy,YU2014121}. In particular, $W_-$ is said to be part of the Wishart ensemble but $W_+$ is said to belong the the Anti-Wishart ensemble. In this naming convention we will specifically study the Anti-Wishart ensemble. Of course, the difference between the two are just the extra degenerate zero eigenvalues in the Anti-Wishart ensemble.}. This is a concrete example of how a random matrix can give rise to both random and non-random sectors. Throughout this section we will be interested in the spectral form factor associated with $W_+$ in the regime where $\Gamma=\tilde{\Gamma}N$, where $\tilde{\Gamma}$ is some order $1$ constant. 

To aid our analysis of the spectral form factor it will be convenient to make use of an object called the ``Christoffel–Darboux kernel'' (or just ``kernel'' for short). The kernel is central in the study of random matrices. Intuitively, it encodes the eigenvalue statistics of the random matrix theory. Generally, the kernel for an ensemble of $N\times N$ random matrices is given by introducing orthogonal polynomials, $P_{k}(x)$, which satisfy the following orthogonality relation:
\begin{equation}
\label{orthRelWishart}
    \int e^{-V(x)}P_k(x)P_\ell(x)dx=\delta_{k\ell}\ ,
\end{equation}
where $V(x)$ is an effective potential which can be derived by an appropriate rewriting of the joint probability density function (JPDF) of eigenvalues. Using these polynomials and the effective potential one then defines the kernel at any finite $N$ as:
\begin{equation}
\label{WishKern}
    K_N(\lambda_i,\lambda_j)=e^{-\frac{1}{2}V(\lambda_i)-\frac{1}{2}V(\lambda_j)}\sum_{p=0}^{N-1}P_{p}(\lambda_i)P_{p}(\lambda_j)\ .
\end{equation}
By construction, the density of states of the matrix ensemble at any finite $N$ is exactly given by $K_N(x,x)$, which satisfies $\int K_N(x,x) dx=N$. The reader can read more details about the various aspects of the kernel discussed above in \cite{forrester2010log,mehta2004random,livan2018introduction} as well as in the brief overview (based on these texts) provided in Appendix \ref{ReviewRMTKernel}.

\begin{figure}[h]
    \centering
    \includegraphics[width=1\linewidth]{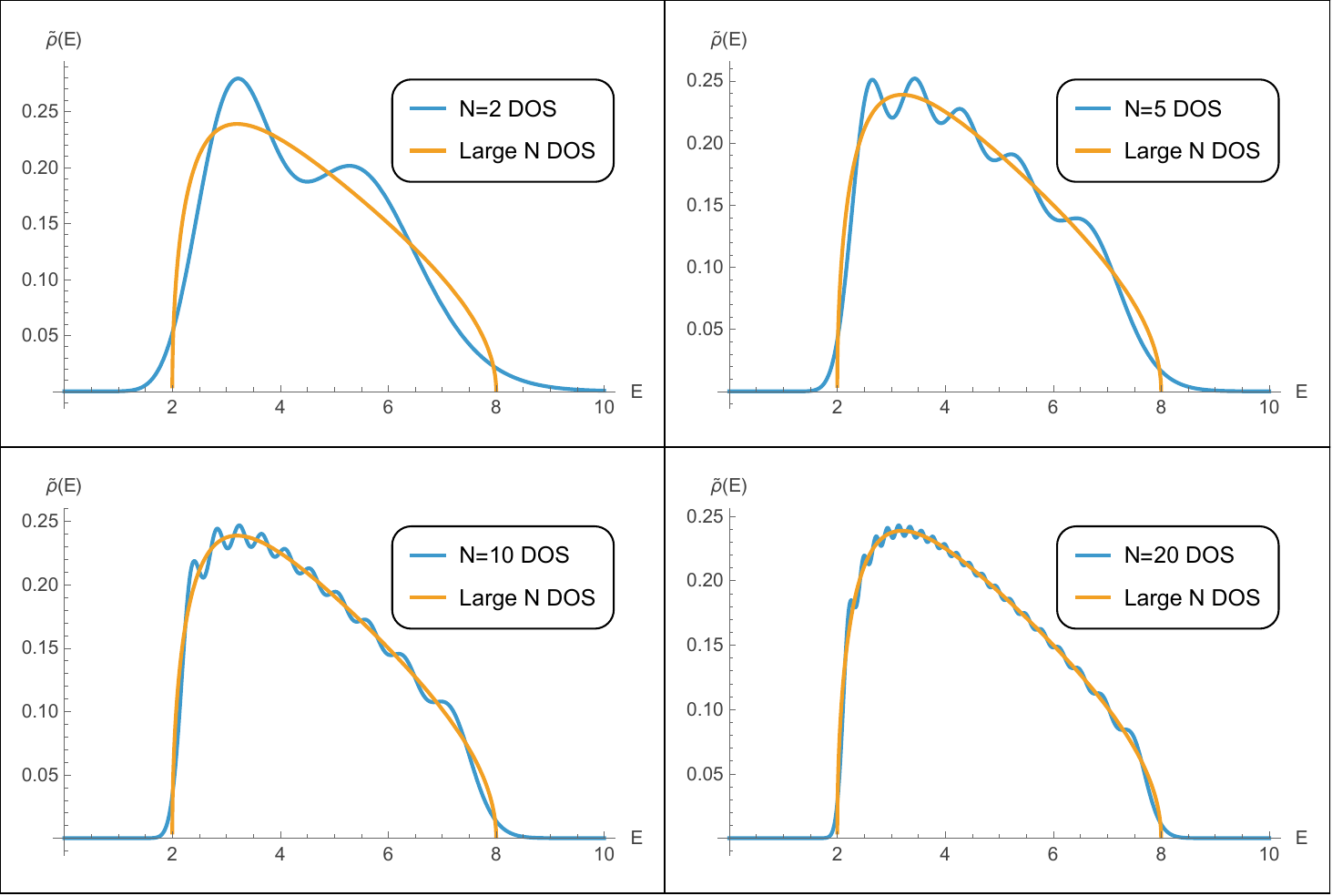}
    \caption{We make plots of the normalized density of states as computed by $K_{sc.}(E,E)$ (blue line) for finite values of $N$ and compare to the Marchenko-Pastur PDF $\tilde{\rho}_{MP}(E)$ (yellow line) for $\tilde{\Gamma}=8$. We can see that as $N$ gets larger the $K_{sc.}(E,E)$ converges towards the Marchenko-Pastur PDF as claimed in Eq. (\ref{MPDistribution}). Note that the plots here only include the random sector the full density will include an extra delta function $\Gamma\delta(E)$ in the plot itself it will be a spike of height $\tilde{\Gamma}$ because we plot $\tilde{\rho}$ (everything divided by $N$).}
    \label{FinteNDOS}
\end{figure}

A key property that we will use later on is that the kernel satisfies the so-called ``reproducing'' property given as:
\begin{equation}
\label{ReproducingPropOfKernel}
    \int K_{N}(x,y)K_{N}(y,x')dy=K_N(x,x')\ ,
\end{equation}
which can be easily checked by plugging in the definition of the kernel in Eq. (\ref{WishKern}) and then evaluating the integral using the orthogonality relation in Eq. (\ref{orthRelWishart}). Another important identity which we will use to simplify the expression of the kernel, is the Christoffel–Darboux formula. It states that for an orthonormal set of polynomials $\{P_n\}_{n=1,..,N-1}$ with the highest order term coefficient given by $k_n$ we have the following identity:
\begin{equation}
    \sum_{n=0}^{N-1}P_n(x)P_n(x')=\frac{k_{N-1}}{k_{N}}\left[\frac{P_{N-1}(x')P_N(x)-P_N(x')P_{N-1}(x)}{x-x'}\right]\ ,
\end{equation}
we also have its so-called confluent form which is given by taking $x'\to x$:
\begin{equation}
    \lim_{x'\to x}\sum_{n=0}^{N-1}P_n(x)P_n(x')=\frac{k_{N-1}}{k_N}\left[P_N'(x)P_{N-1}'(x)-P_{N-1}'(x)P_N(x)\right]\ .
\end{equation}
In the specific case of the complex Wishart ensemble we have the effective potential:
\begin{equation}
    V(x)=x-\Gamma\ln(x)\ .
\end{equation}
For this effective potential, the appropriate choice of polynomials that satisfy Eq. (\ref{orthRelWishart}) will be the generalized Laguerre polynomials\footnote{For this reason Wishart ensembles are also referred to as Wishart-Laguerre ensembles.} denoted $L_n^{(\Gamma)}(x)$, properly normalized we have:
\begin{equation}
    P_n^{(\Gamma)}(\lambda)=\frac{L_n^{(\Gamma)}(\lambda)}{\sqrt{(\Gamma+n)!/n!}}\ ,
\end{equation}
so that they satisfy:
\begin{equation}
    \int_0^\infty e^{-\lambda+\Gamma\ln(\lambda)}P_n^{(\Gamma)}(\lambda)P_m^{(\Gamma)}(\lambda) d\lambda=\delta_{nm}\ .
\end{equation}
Using the Christoffel–Darboux formula we obtain the following explicit expressions for the kernel \footnote{In the case of the generalized Laguerre polynomials the $k_n$ in the Christoffel–Darboux formula is given as $k_n=\frac{(-1)^n}{\sqrt{n!(\Gamma+n)!}}$.}:
\begin{equation}
\begin{split}
    &K_N(x_1,x_2)=-\sqrt{N(N+\Gamma)}x_1^{\frac{\Gamma}{2}}x_2^{\frac{\Gamma}{2}}e^{-\frac{1}{2}(x_1+x_2)}\left[\frac{P^{(\Gamma)}_{N-1}(x_1)P^{(\Gamma)}_N(x_2)-P^{(\Gamma)}_N(x_1)P^{(\Gamma)}_{N-1}(x_2)}{x_2-x_1}\right]\\
    &=\frac{N!}{(N+\Gamma-1)!}x_1^{\frac{\Gamma}{2}}x_2^{\frac{\Gamma}{2}}e^{-\frac{1}{2}(x_1+x_2)}\left[\frac{L^{(\Gamma)}_{N-1}(x_1)L^{(\Gamma)}_N(x_2)-L^{(\Gamma)}_N(x_1)L^{(\Gamma)}_{N-1}(x_2)}{x_1-x_2}\right]\ ,\\
\end{split}
\end{equation}
and density of states of the complex Wishart ensemble
\begin{equation}
    \begin{split}
        & K_N(x,x)=\rho(x)=-\sqrt{N(N+\Gamma)}x^{\Gamma}e^{-x}\left[P_{N-1}^{(\Gamma)}(x)\frac{d}{dx}P_{N}^{(\Gamma)}(x)-P_{N}^{(\Gamma)}(x)\frac{d}{dx}P_{N-1}^{(\Gamma)}(x)\right]\\
    &=\frac{N!}{(N+\Gamma-1)!}x^{\Gamma}e^{-x}\left[L_{N-1}^{(\Gamma)}(x)L_{N-1}^{(\Gamma+1)}(x)-L_{N}^{(\Gamma)}(x)L_{N-2}^{(\Gamma+1)}(x)\right]\ .\\
    \end{split}
\end{equation}
It is possible to recover the well known Marchenko–Pastur distribution in the large $N$ regime by using the expression for the kernel in a certain scaling limit \cite{marchenko1967distribution,livan2018introduction}. We consider a scaled kernel denoted $K_{sc.}(x,x')$ which we define as:
\begin{equation}
    K_{sc.}(E,E')=2K_N(2NE,2NE')\ .
\end{equation}
The purpose of the scaling in the eigenvalues is to ensure that the ``endpoints'' of the distribution remained fixed as we increase $N$ which is a standard procedure when discussing the large $N$ regime in random matrix theory\footnote{Note that depending on the ensemble the scaling with $N$ will be different. For example for the GUE the set of polynomials are the Hermite polynomials and the scaling will go as $\sqrt{N}$ \cite{livan2018introduction}.}. The Marchenko–Pastur PDF appears as the infinite $N$ limit of the diagonal of the scaled kernel \cite{marchenko1967distribution,livan2018introduction,Johnson:2024tgg}\footnote{The paper \cite{Liu:2018hlr} studied the form factor in the case where $\tilde{\Gamma}=0$ whereas we will generally be interested in the regime where $\tilde{\Gamma}\geq 1$.}:
\begin{equation}
\label{MPDistribution}
\begin{split}
    &\lim_{N\to\infty}K_{sc.}(E,E)=\tilde{\rho}_{MP}(E)=\left[\frac{1}{\pi}\frac{\sqrt{(E_+-E)(E-E_-)}}{E}\right]\\
    &E_{\pm}=\frac{1}{2}\left(1\pm\sqrt{\tilde{\Gamma}+1}\right)^2\ .\\
\end{split}
\end{equation}
In Figure \ref{FinteNDOS}, we illustrate the exact spectral density for various increasing values of $N$ and compare how close it is to the Marchenko–Pastur distribution. 
We can see that as we increase the value of $N$ the distribution converges to the Marchenko–Pastur distribution, as expected. We can also clearly see the emergence of the sharp macroscopic gap, of size~$E_{Gap}=E_-$, between the degenerate states at $E=0$ and the excited continuum of states described  by the  Marchenko–Pastur distribution.

By utilizing the diagonal of the kernel one can write the disconnected form factor of the Wishart ensemble (including the presence of the degenerate states) as:
\begin{equation}
\label{DisconnFFKernFormula}
\begin{split}
     &\braket{Z(\beta+it)Z(\beta-it)}_{\text{dis.}}=N^2\left[\tilde{\Gamma}+\int_0^\infty\tilde{\rho}(E)e^{-(\beta+it)E} dE\right]\left[\tilde{\Gamma}+\int_0^\infty\tilde{\rho}(E)e^{-(\beta-it)E} dE\right]\ ,\\
\end{split}
\end{equation}
with $\tilde{\rho}(E)=K_{sc.}(E,E)$ and at large $N$ we can replace $\tilde{\rho}=\tilde{\rho}_{MP}$. 

The connected part of the spectral form factor can also be expressed in terms of the kernel and is given by \cite{Johnson:2020exp,Johnson:2020mwi,Johnson:2022wsr}:
\begin{equation}
\label{ConFFKernelFormula}
    \begin{split}
        &\braket{Z(\beta+it)Z(\beta-it)}_{\text{con.}}\\
        &=\braket{Z_R(2\beta)}-N^2\int_0^\infty dE_1\int_0^\infty dE_2 K_{sc.}(E_1,E_2)K_{sc.}(E_2,E_1)e^{-\beta(E_1+E_2)}e^{-it(E_1-E_2)}\\
        &\braket{Z_R(2\beta)}=N\int_0^\infty dE K_{sc.}(E,E)e^{-2\beta E}=N\int_0^\infty dE \tilde{\rho}(E)e^{-2\beta E}\ .\\
    \end{split}
\end{equation}
The extra term, $\braket{Z_R(2\beta)}$, is required to ensure the value of the plateau of the connected part matches the results in Table \ref{FFValuesWithDegens}. Together Eq. (\ref{DisconnFFKernFormula}) and Eq. (\ref{ConFFKernelFormula}) give the prescription to compute the spectral form factor in terms of the kernel and explicitly include the effects of the degenerate ground states. Using these expressions we will proceed to study the ensemble averaged form factor using both numerical and analytic computations.

\subsection{Disconnected Part of Spectral Form Factor}
Recall that to obtain the disconnected spectral form factor we can use Eq. (\ref{DisconnFFKernFormula}) which, for the complex Wishart ensemble, will explicitly read:
\begin{equation}
    \begin{split}
        &\braket{Z_R(\beta)}=N\int_0^\infty K_{sc.}(E,E)e^{-\beta E} dE=2N\int_0^\infty K_N(2NE,2NE)e^{-\beta E}dE=\int_0^\infty K_{N}(x,x)e^{-\frac{\beta x}{2N}}dx\\
        &=\frac{N!}{(N+\Gamma-1)!}\int_0^\infty e^{-\frac{\beta x}{2N}}e^{-x}x^{\Gamma}\left[L_{N-1}^{(\Gamma)}(x)L_{N-1}^{(\Gamma+1)}(x)-L_{N}^{(\Gamma)}(x)L_{N-2}^{(\Gamma+1)}(x)\right]dx\ .\\
    \end{split}
\end{equation}
Such an integral can be evaluated exactly for any finite $N$. This is due to the fact that we are dealing with a polynomial integrated against an exponential function in the measure. In particular, the combination of Laguerre polynomials in the square brackets can be expressed as:
\begin{equation}
    \begin{split}
    \label{DefiitionofCJ}
        &L_{N-1}^{(\Gamma)}(x)L_{N-1}^{(\Gamma+1)}(x)-L_{N}^{(\Gamma)}(x)L_{N-2}^{(\Gamma+1)}(x)=\sum_{J=0}^{2N-2} C_J(N,\Gamma) x^J\ ,\\
    \end{split}
\end{equation}
where $C_J$ is a constant that depends on $\Gamma$ and $N$. Using such an expansion gives:
\begin{equation}
\begin{split}
    &\braket{Z_R(\beta)}=\frac{N!}{(N+\Gamma-1)!}\int_0^\infty e^{-\left(1+\frac{\beta}{2N}\right)x}\sum_{J=0}^{2N-2} C_J(N,\Gamma)x^{J+\Gamma}dx\\
    &=\frac{N!}{(N+\Gamma-1)!}\sum_{J=0}^{2N-2}C_J(N,\Gamma)\left[\int_0^\infty e^{-(1+\frac{\beta}{2N})x}x^{J+\Gamma} dx\right]\\
    &=\frac{N!}{(N+\Gamma-1)!}\sum_{J=0}^{2N-2} C_J(N,\Gamma) \frac{(J+\Gamma)!}{\left(1+\frac{\beta}{2N}\right)^{J+\Gamma+1}}\ .\\
\end{split}
\end{equation}
To completely define the full exact result we need to give an expression for $C_J$. This is done in work detailed in Appendix \ref{DeriveFormulaCJAppendix} the final result is:
\begin{equation}
\label{FormulaForCJ}
\begin{split}
    &C_J(N,\Gamma)=\frac{(-1)^J}{J!}\left[\binom{\Gamma+N-1}{N-J-1}\binom{N+\Gamma}{N-1}f_{J,N,\Gamma}(1)-\binom{\Gamma+N-1}{N-J-2}\binom{N+\Gamma}{N}f_{J,N,\Gamma}(0)\right]\\
    &f_{J,N,\Gamma}(n)={}_3F_2\left(-J,n-N,n-1-J-\Gamma;N-J-1+n,\Gamma+1+n;-1\right)\ .\\
\end{split}  
\end{equation}
With this, we have the following expression for the exact finite $N$ partition function of the non-degenerate sector:
\begin{equation}
    \begin{split}
        &\braket{Z_R(\beta)}=\sum_{J=0}^{2N-2}\frac{c_J(N,\Gamma)}{\left(1+\frac{\beta}{2N}\right)^{J+\Gamma+1}}\\
        &c_J(N,\Gamma)=(-1)^J\frac{(J+\Gamma)!N!}{(N+\Gamma-1)!}C_J(N,\Gamma)\ .\\
    \end{split}
\end{equation}

Although the expression we derived could in principle be used to exactly understand the disconnected part of the form factor it would be a relatively complicated to extract the physics from such a result. Instead, what we will opt for is to go to a regime of sufficiently large $N$ and take $\tilde{\rho}(E)=\tilde{\rho}_{MP}$. Then the partition function can be expressed as:
\begin{equation}
\begin{split}
    &\braket{Z(\beta)}=N\tilde{\Gamma}+N\int_0^\infty \tilde{\rho}(E)e^{-\beta E}dE=N\left[\tilde{\Gamma}+\int_{E_-}^{E_+} \frac{\sqrt{(E_+-E)(E-E_-)}}{\pi E}e^{-\beta E}dE\right]\ .\\
\end{split}
\end{equation}
We can expand the $e^{-\beta E}=1+\sum_{k=1}^\infty \frac{(-\beta)^k}{k!}E^k$ and then integrate term by term to get:
\begin{equation}
    \begin{split}
        &\braket{Z(\beta)}=\Gamma+N\int_{E_-}^{E_+}\frac{\sqrt{(E_+-E)(E-E_-)}}{\pi E}\left(1+\sum_{k=1}^\infty \frac{(-\beta)^k}{k!}E^k\right)dE\\
        &=\Gamma+N+N\sum_{k=1}^\infty \frac{(-\beta)^k}{k!}\int_{E_-}^{E_+}\frac{\sqrt{(E_+-E)(E-E_-)}}{\pi E}E^{k}dE\\
        &=\Gamma+N+N\sum_{k=1}^\infty \frac{(-\beta)^k}{k!}\left(\frac{E_-^{k-1}(E_+-E_-)^2}{8}{}_2F_1\left[\frac{3}{2},1-k;3;1-\frac{E_+}{E_-}\right]\right)\ .\\
    \end{split}
\end{equation}
Note that $E_{\pm}$ depend on $\tilde{\Gamma}$. Now consider the limit where $\tilde{\Gamma}\gg 1$ (i.e. the gap size is very large) and find:
\begin{equation}
    \begin{split}
        &\braket{Z(\beta)}=\Gamma+N+N\sum_{k=1}^\infty \frac{(-\beta)^k}{k!}\tilde{\Gamma}^k\left[\sum_{p=0}^\infty \frac{2^{-k}(k)_{p+1}\tilde{\Gamma}^{-p}}{(1)_p(2)_p(k)_{1-p}}\right]\\
        &=N\left[\tilde{\Gamma}+1+\sum_{k=1}^\infty \sum_{p=0}^k \left[\frac{\left(-\beta\tilde{\Gamma}/2\right)^k}{k!} \frac{(k+p)!}{p!(p+1)!(k-p)!}\right]\tilde{\Gamma}^{-p}\right]\\
        &=N\left[\tilde{\Gamma}+e^{-\beta\tilde{\Gamma}/2}\left(1+f(\beta,\tilde{\Gamma})\right)\right]\\
        &f(\beta,\tilde{\Gamma})=e^{\beta \tilde{\Gamma}/2}\sum_{k=1}^\infty \sum_{p=1}^k \left[\frac{\left(-\beta\tilde{\Gamma}/2\right)^k}{k!} \frac{(k+p)!}{p!(p+1)!(k-p)!}\right]\tilde{\Gamma}^{-p}\ ,\\
    \end{split}
\end{equation}
where $(k)_p=\text{Pochhammer}[k,p]$. We see that ignoring the $f(\beta,\tilde{\Gamma})$ and computing the disconnected spectral form factor results in a purely oscillating function whose period is given as $\tau=\frac{4\pi}{\tilde{\Gamma}}$. In the very large $\tilde{\Gamma}$ limit we have $E_-=E_{gap}=\frac{\tilde{\Gamma}}{2}$, and we can reinterpret the expression for the period in terms of the gap size and write $\tau=\frac{2\pi}{E_{gap}}$. When we compare the width of the density of the non-degenerate sector, which goes as $E_+-E_-\sim 2\sqrt{\tilde{\Gamma}}$, and the size of the gap, scaling as $E_-=\tilde{\Gamma}$, we can clearly see that the continuum density is effectively becoming highly localized a distance $E_{gap}$ away from the degenerate states. Based on this observation one can give a crude estimate for the resulting density which will be:
\begin{equation}
    \rho(E)=N\left[\tilde{\Gamma}\delta(E)+\delta\left(E-\frac{\tilde{\Gamma}}{2}\right)\right]\ ,
\end{equation}
which will exactly reproduce the expression for the disconnected form factor we got by ignoring the contribution from $f(\beta,\tilde{\Gamma})$.

\begin{figure}[]
    \centering
    \includegraphics[width=0.9\linewidth]{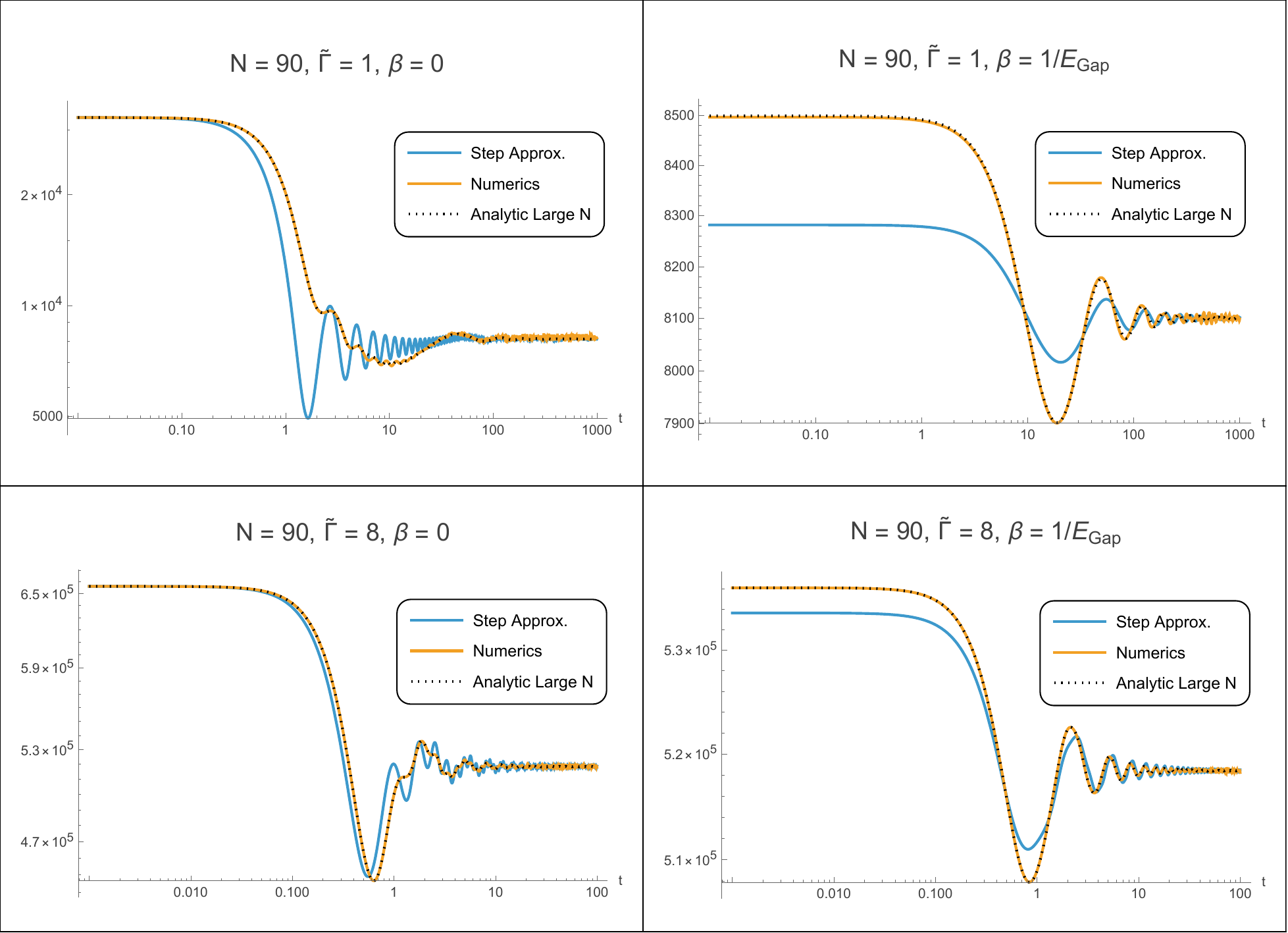}
    \caption{We plot the spectral form factor for the Wishart ensemble at $N=90$ at infinite (top and bottom left plots) and at finite temperature $T=E_{Gap}=E_-$ (top and bottom right plots) for various values of $\tilde{\Gamma}$. The orange line is the numeric computation of the ensemble averaged form factor generated by actually computing the averaged form factor from computer simulated eigenvalues over many trials. The dotted line is the computation of the disconnected part of the form factor which is obtained by numerically computing the integral given in Eq. (\ref{DisconnFFKernFormula}). The blue line is the spectral form factor associated with step approximation given by the expression in Eq. (\ref{StepApproxSFF}).}
    \label{NumericVsAnalyticDisconMPFF}
\end{figure}

Another more refined approximation which we can consider which will result in damped oscillations comes from replacing the delta function with a step of width $E_+-E_-$. In this case, we will get the following expression for the disconnected form factor:
\begin{equation}
\label{StepApproxSFF}
    \begin{split}
     &\frac{\braket{Z(\beta+it)Z(\beta-it)}_{dis.}}{N^2}=\tilde{\Gamma}^2\\
     &+\frac{2\tilde{\Gamma}\left[e^{-\beta E_-}\left(\beta \cos(E_-t)-t\sin(E_-t)\right)+e^{-\beta E_+}\left(-\beta \cos(E_+t)+t\sin(E_+t)\right)\right]}{(E_+-E_-)(t^2+\beta^2)}\\
     &-\frac{2e^{-\beta(E_++E_-)}\left[\cos\left((E_+-E_-)t\right)-\cosh\left((E_+-E_-)\beta\right)\right]}{(E_+-E_-)^2(t^2+\beta^2)}\ .\\
    \end{split}
\end{equation}
We can see that when $\beta\neq 0$ the leading time dependent terms will be the ones multiped by $e^{-\beta E_-}$ (due to the following hierarchy $0\leq E_-<E_+<E_++E_-$). So we have a rough estimate for the time dependence of the disconnected form factor given as:
\begin{equation}
    \begin{split}
     &\frac{\braket{Z(\beta+it)Z(\beta-it)}_{dis.}}{N^2}\approx \tilde{\Gamma}^2+\frac{2\tilde{\Gamma}\left[e^{-\beta E_-}\left(\beta \cos(E_-t)-t\sin(E_-t)\right)\right]}{(E_+-E_-)(t^2+\beta^2)}+\cdot\cdot\cdot\ .\\
    \end{split}
\end{equation}
We can clearly see that at late enough times there will be oscillations with a period, $\tau$, well approximated by the size of the gap $E_{Gap}=E_-$:
\begin{equation}
    \tau \approx \frac{2\pi}{E_-}\ .
\end{equation}
In Figure \ref{NumericVsAnalyticDisconMPFF} we compare the results of the step approximation we have done to actual numeric simulations of the full form factor as well as the large $N$ computations for the disconnected part for $N=90$ at various values of $\beta$ and $\tilde{\Gamma}$.

\begin{figure}[]
    \centering
    \includegraphics[width=0.9\linewidth]{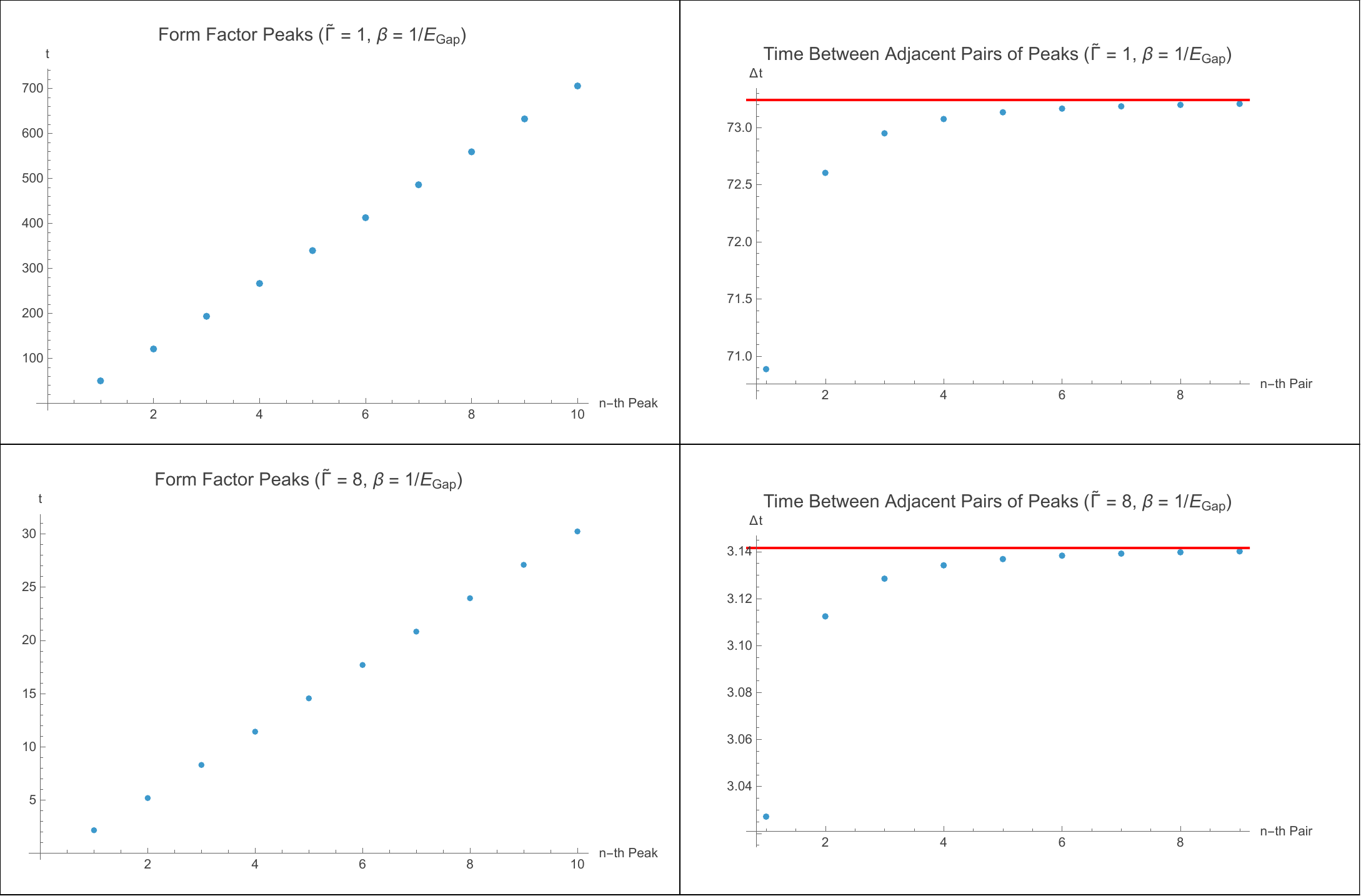}
    \caption{The two left panels are plotting the time at which the $n-th$ peaks in the oscillations appear for the in the form factor appear at a temperature scale set by $E_{Gap}=E_-$ for different values of $\tilde{\Gamma}$. Next to them on the right side of the figure we compute the time difference between the $n$-th adjacent pair of peaks. We can see that the time difference between peaks approaches the value $\tau=\frac{2\pi}{E_{Gap}}=\frac{2\pi}{E_-}$ whose numerical value is represented by the red horizontal line.}
    \label{FFMPPeriodPlot}
\end{figure}

As we can see, the numeric calculation of the total form factor (i.e. is the sum of the connected and disconnected parts) is captured to astounding accuracy by simply computing the disconnected part of the form factor using the large $N$ analytic density of states. This clearly demonstrates what we alluded to in Section \ref{GenAspFFSection}, namely that the connected part of the form factor will always represent a sub-leading contribution at all time scales. We see that at infinite temperature we have complicated oscillations (``beats'') in the form factor towards the plateau but as we lower the temperature to be of the order $E_{\text{Gap}}=E_-$ the structure of the oscillations become more simple to see and analyze. It is in such a regime, we expect the time between the oscillation peaks of the form factor to be characterized by the gap. This is verified by the sample plots in Figure \ref{FFMPPeriodPlot}. We can see the time between peaks (plots on the left) appear to form a straight with a slope that is very well approximated by~$2\pi/E_-$. In fact, a more refined analysis measuring the time delay between adjacent peaks (plots on the right) show that the period of oscillations are exactly approaching the value $2\pi/E_-$ indicated by the horizontal red line.

Based on this analysis we conclude that at sufficiently low temperatures (of the order of $E_{Gap}=E_-$) the presence of the degenerate eigenvalues imprint on the disconnected form factor as damped oscillations whose period is set by the inverse size of the gap.

\newpage
\subsection{Connected Part of Spectral Form Factor}
\begin{figure}[h!]
    \centering
    \includegraphics[width=1\linewidth]{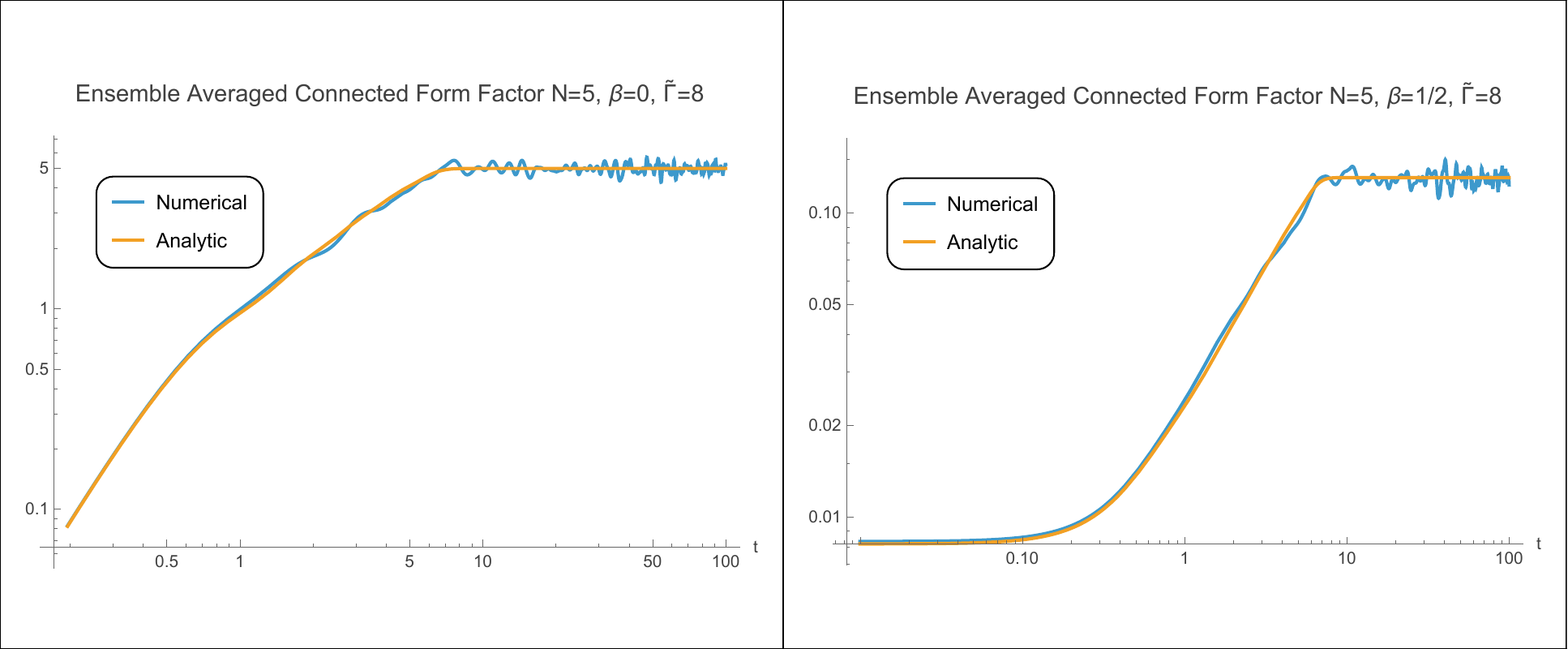}
    \caption{We compare computations of the form factor via the expression given by the analysis of the kernel (orange line) given in Eq. (\ref{WishartConnFiniteNSFFAnalytic}) to numeric computations given by computer generated eigenvalues averaged over many trials (blue line). We can see very close agreement. Deviations of the blue line from the yellow line are due to the finite number of trials used to generate the blue line.}
    \label{N5NumVsAnalyticConFF}
\end{figure}

Here we will address the computation of the sub-leading connected part of the spectral form factor given by:
\begin{equation}
    \begin{split}
        &\braket{Z(\beta+it)Z(\beta-it)}_{\text{con.}}\\
        &=\braket{Z_R(2\beta)}-N^2\int_0^\infty dE_1\int_0^\infty dE_2 K_{sc.}(E_1,E_2)K_{sc.}(E_2,E_1)e^{-\beta(E_1+E_2)}e^{-it(E_1-E_2)}\\
        &\braket{Z_R(2\beta)}=N\int_0^\infty dE K_{sc.}(E,E)e^{-2\beta E}=N\int_0^\infty dE \tilde{\rho}(E)e^{-2\beta E}\ .\\
    \end{split}
\end{equation}
We will start our analysis by simply plugging in the expression for the kernel into the integral and access how far we can get (at least in principle) towards exactly computing the connected part. We are interested in computing the following integral:
\begin{equation}
\begin{split}
    &\mathcal{I}_N(\beta_1,\beta_2)=-N^2\int_0^\infty dE_1\int_0^\infty dE_2 K_{sc.}(E_1,E_2)K_{sc.}(E_2,E_1)e^{-\beta_1 E_1}e^{-\beta_2 E_2}\\
    &=-4N^2\int_0^\infty dE_1\int_0^\infty dE_2 K_{N}(2NE_1,2NE_2)K_{N}(2NE_2,2NE_1)e^{-\beta_1 E_1}e^{-\beta_2 E_2}\\
    &=-\int_0^\infty dx_1 \int_0^\infty dx_2\left[K_N(x_1,x_2)\right]^2e^{-\frac{\beta_1x_1}{2N}}e^{-\frac{\beta_2x_2}{2N}}\\
    &=-\mathcal{C}_N^2\int_0^\infty dx_1 \int_0^\infty dx_2 x_1^\Gamma x_2^\Gamma\left[\frac{L^{(\Gamma)}_{N-1}(x_1)L^{(\Gamma)}_N(x_2)-L^{(\Gamma)}_N(x_1)L^{(\Gamma)}_{N-1}(x_2)}{(x_1-x_2)}\right]^2e^{-\left(1+\frac{\beta_1}{2N}\right)x_1}e^{-\left(1+\frac{\beta_2}{2N}\right)x_2}\\
    &\mathcal{C}_N=\frac{N!}{(N+\Gamma-1)!}\ .\\
\end{split}
\end{equation}
We can consider the Taylor expansion of the combination of Laguerre polynomials in the square brackets in powers of $(x_1-x_2)$: 
\begin{equation}
\begin{split}
    &L^{(\Gamma)}_{N-1}(x_1)L^{(\Gamma)}_N(x_2)-L^{(\Gamma)}_N(x_1)L^{(\Gamma)}_{N-1}(x_2)\\
    &=\sum_{n=1}^{\infty} \frac{1}{n!}\frac{\partial^n}{\partial x_1^n}\left[L_{N-1}^{(\Gamma)}(x_1)L_N^{(\Gamma)}(x_2)-L_N^{(\Gamma)}(x_1)L_{N-1}^{(\Gamma)}(x_2)\right]\bigg\vert_{x_1\to x_2}(x_1-x_2)^n\\
    &=\sum_{n=1}^N b_n(N,\Gamma,x_2)(x_1-x_2)^n\\
    &b_n=b_n(N,\Gamma,x_2)=\frac{(-1)^n}{n!}\left[L_{N-n-1}^{(\Gamma+n)}(x_2)L_N^{(\Gamma)}(x_2)-L_{N-n}^{(\Gamma+n)}(x_2)L_{N-1}^{(\Gamma)}(x_2)\right]\ .\\
\end{split}
\end{equation}
The series expansion truncates to a finite polynomial of degree $N$ in $x_1-x_2$ due to the fact that $\frac{d^k}{dx^k}L_n^{(\alpha)}(x)=0$ when $k>n$. The coefficients $b_n(N,\Gamma,x_2)$ themselves are polynomials in $x_2$. From this it follows that: 
\begin{equation}
\label{WishartConnFiniteNSFFAnalytic}
    \begin{split}
        &\mathcal{I}_N(\beta_1,\beta_2)=-\mathcal{C}_N^2\int_0^\infty dx_1 \int_0^\infty dx_2x_1^\Gamma x_2^\Gamma e^{-\left(1+\frac{\beta_1}{2N}\right)x_1} e^{-\left(1+\frac{\beta_2}{2N}\right)x_2} \left[\sum_{n=1}^Nb_n(x_1-x_2)^{n-1}\right]^2\\
        &\Rightarrow \braket{Z(\beta+it)Z(\beta-it)}_{con.}=\braket{Z_R(2\beta)}+\mathcal{I}_N(\beta+it,\beta-it)\ .\\
    \end{split}
\end{equation}
In this form, it is completely clear for any given finite $N$ it is, in principle, possible to get an exact result. This is because by expanding out the terms in the square brackets we will obtain terms involving products of $x_1$ and $x_2$ which can be integrated term by term exactly\footnote{We also refer the reader to \cite{Forrester_2021} for a more mathematically precise treatment of the connected form factor of Wishart ensembles.}. Even for relatively small values of $N$ the number of terms involved becomes quite large and although the results are exact it is difficult to extract physics from such a result. Nonetheless, the exact result gives us something to compare our approximations and numerical computations to. In particular, in Figure \ref{N5NumVsAnalyticConFF} we give a few plots of the exact connected form factor for $N=5$ and compare to numerical simulations of the averaged form factor (averaged over 300 samples) and find agreement between the two approaches.

Thus far, we have discussed the connected form factor in terms of the exact formulation in terms of orthogonal polynomials. We found that although one can in principle compute things exactly for any $N$ the physics is obscured due to the sheer number of terms involved. The goal of the analysis to follow is to provide a more coarse grained way to analyze the connected part of the form factor similar in sprit to the analysis we did for the disconnected form factor. Although we will sacrifice precision in such computations we will gain more physical insight into what role various parameters defining our system has on the connected part of the spectral form factor. 

\begin{figure}[h!]
    \centering
    \includegraphics[width=0.8\linewidth]{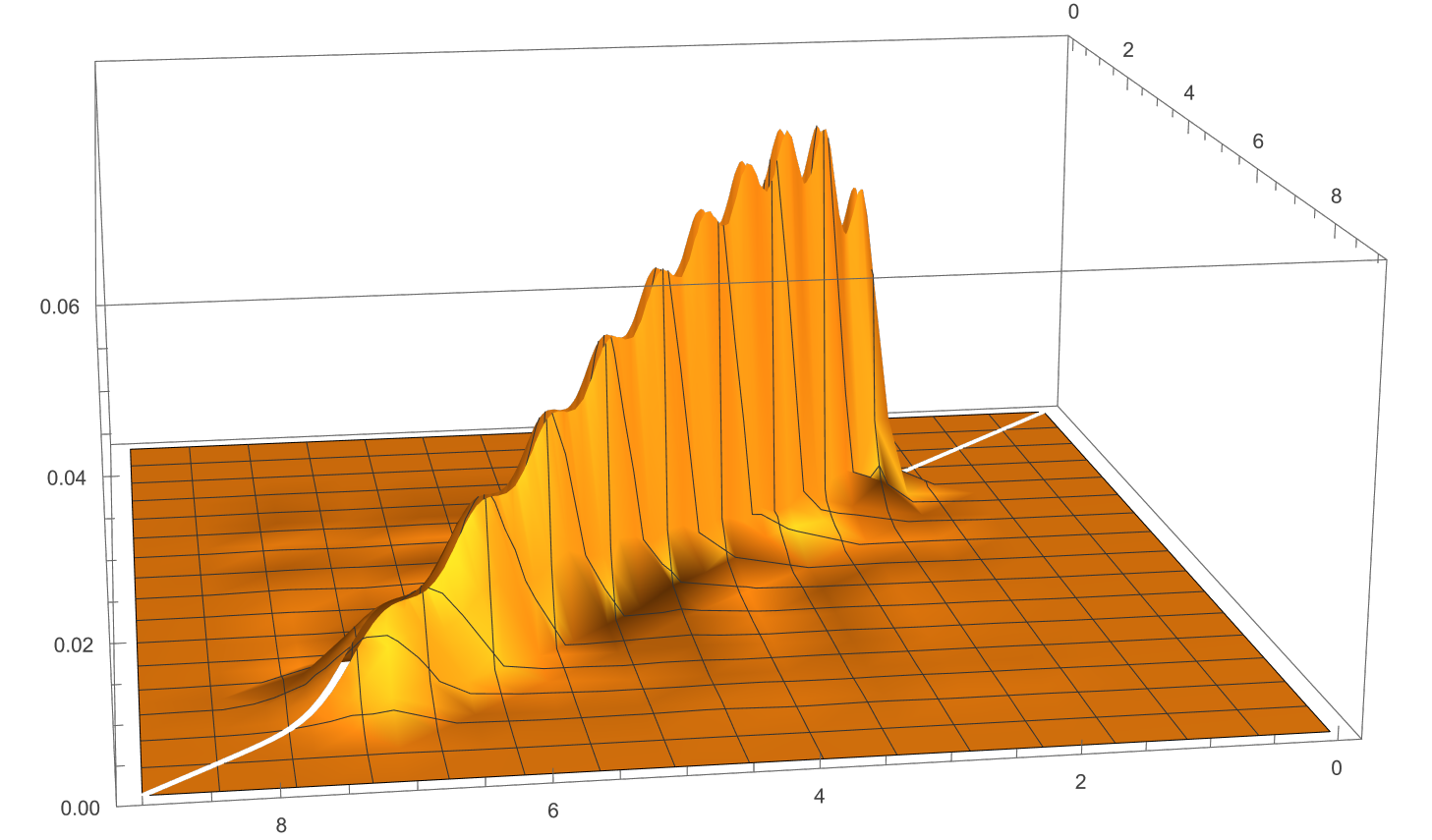}
    \caption{A 3D plot of $K_{sc.}(x_1,x_2)^2$ for $N=10,\tilde{\Gamma}=8$.}
    \label{KerSqWishart}
\end{figure}

To begin, it is useful to to plot $K_{sc.}(x_1,x_2)^2$ since it is the main input of the integrand that defines the connected part of the spectral form factor. In Figure \ref{KerSqWishart}, we make a sample plot of the kernel squared for $N=10$ and $\tilde{\Gamma}=8$.
As was noted in \cite{Johnson:2022wsr}, we can see the kernel squared looks like a ``fin'', it runs along the $x_1=x_2$ plane starting around $E_-$ and ending around $E_+$. Based on such observations it is natural to adopt the following change of variables:
\begin{equation}
\begin{split}
    &x_1=x_+-x_-, \qquad x_2=x_++x_-\ .\\
\end{split}
\end{equation}
Since $x_1,x_2>0$, it follows that $x_+>|x_-|$. We define, $\tilde{K}(x_+,x_-)=K_{sc.}(x_+-x_-,x_++x_-)$ and write:
\begin{equation}
    \begin{split}
        \braket{Z(\beta+it)Z(\beta-it)}_{con.}=\braket{Z_R(2\beta)}-N^2\left[2\int_{0}^\infty dx_+e^{-2\beta x_+}\int_{-x_+}^{x_+}dx_-e^{i2tx_-}\tilde{K}(x_+,x_-)\tilde{K}(x_+,-x_-)\right]\ .
    \end{split}
\end{equation}
The statement that $K_{sc.}(x_1,x_2)$ is symmetric under the interchange of $x_1$ and $x_2$ translates to the statement that $\tilde{K}(x_+,x_-)$ is an even function in $x_-$, for any choice of $x_+>0$. Using this we can write:
\begin{equation}
\begin{split}
    &\braket{Z(\beta+it)Z(\beta-it)}_{con.}=\braket{Z_R(2\beta)}-4N^2\int_{0}^\infty dx_+e^{-2\beta x_+}\int_{0}^{x_+}dx_-\cos(2tx_-)\tilde{K}(x_+,x_-)^2\ .\\
\end{split}
\end{equation}
The key feature we want to capture in our approximations of the kernel is how quickly the ``fin'' decays as we move away from $|x_-|=0$. Of course, this will depend on $x_+$ but we will ignore this fact and approximate the kernel to be of the form:
\begin{equation}
    \tilde{K}(x_+,x_-)=h(x_-)\tilde{K}(x_+,0)\ ,
\end{equation}
where $h(x_-)$ is a uniform decay envelope which we are free to choose under the constraints that:
\begin{equation}
\label{ConstraintsForApprox}
    \begin{split}
        &\int_0^\infty dx_+ \tilde{K}(x_+,0)^2 \int_0^{x_+} dx_- h(x_-)^2=\frac{1}{4N}\\
        &h(0)=1\ ,\\
    \end{split}
\end{equation}
The condition that $h(0)=1$ ensures that we reproduce the properly normalized spectral density of when restricting to the diagonal of the kernel at $x_-=0$. The condition involving an integral equaling $(4N)^{-1}$ is a condition that one can arrive at by using the reproducing property of the kernel given in Eq. (\ref{ReproducingPropOfKernel}). In particular, this condition fixes the value of the connected form factor to be zero at infinite temperature at time $t=0$. Now that we have defined the conditions we want to use to define the approximation we will consider two particular decay envelopes and study what they tell us about how the connected form factor depends on $\tilde{\Gamma}$. The first will be a Gaussian decay and the second will correspond to a matrix model inspired sine-kernel decay.\\

\noindent \textbf{Gaussian Approximation for Kernel:}

\noindent For the Gaussian approximation we write the kernel as:
\begin{equation}
    \begin{split}
        &\tilde{K}(x_+,x_-)= e^{\frac{-x_-^2}{4\sigma^2}}\tilde{K}(x_+,0)\ .\\
    \end{split}
\end{equation}
By inspection, we can see for any value of $\sigma>0$ we will correctly reproduce the diagonal of the kernel (i.e. $h(0)=1$ condition is satisfied). All that is left is to fix $\sigma$ which will be done through the integral condition in Eq. (\ref{ConstraintsForApprox}). With a Gaussian decay envelope the condition will read:
\begin{equation}
    \int_0^\infty dx_+\tilde{K}(x_+,0)^2\int_0^{x_+}dx_-e^{-\frac{x_-^2}{2\sigma^2}}=\frac{1}{4N}\ .
\end{equation}
As we can see the integral over $x_-$ must be done first and the result will be a function of~$x_+$ which then integrated at the end. Doing this integral exactly and then inverting the expressions for $\sigma$ will generally be difficult. However, in a certain approximation which we simply refer to as the ``decoupling'' limit/approximation we can find simple expressions. The ``decoupling'' approximation effectively changes the upper limit in the integral with respect to the variable $x_-$ from $x_+$ to infinity. The approximation assumes that $\sigma$ is sufficiently small such that the decay of the fin is very fast so that any errors that are introduced in extending the limits of integration from $x_+$ to infinity are ``small''\footnote{It should also be noted that we should also require the $\sigma\ll E_-$ if this condition fails then the density could be quite large near $x_+,x_-=0$ in which case we really should keep the $x_+$ bound on the first integral with respect to $x_-$. In our discussions we will assume we are away from such a problematic regimes and consider $\tilde{\Gamma}\geq 1$.}. Once we do this we can see the double integral ``decouples'' we can do each integral independently of the other. For the Gaussian approximation in the decoupled approximation we find the following condition on~$\sigma$:
\begin{equation}
    \sigma\sqrt{\frac{\pi}{2}}\int_0^\infty dx_+\tilde{K}(x_+,0)^2=\frac{1}{4N}\ .
\end{equation}
Solving for~$\sigma$ gives~$\sigma=\alpha N^{-1}$ where $\alpha$ is given by:
\begin{equation}
    \alpha= \frac{\left[\int_0^\infty dx_+ \tilde{K}(x_+,0)^2\right]^{-1}}{2\sqrt{2\pi}}=\frac{\left[\int_0^\infty dx_+ \tilde{\rho}(x_+)^2\right]^{-1}}{2\sqrt{2\pi}}\ .
\end{equation}
In the regime where $N\to\infty$ we can evaluate the integral exactly by noting that the density of states approaches the Marchenko-Pastur distribution and obtain:
\begin{equation}
\label{alphaCriticalMP}
    \alpha=\alpha_{c}=\frac{1}{2\sqrt{2\pi}\int_0^\infty dx_+ \tilde{\rho}_{MP}(x_+)^2}=\frac{\pi^{3/2}}{2\sqrt{2}\left[2\left(\tilde{\Gamma}+2\right)\text{arctanh}\left(\frac{2\sqrt{1+\tilde{\Gamma}}}{2+\tilde{\Gamma}}\right)-4\sqrt{1+\tilde{\Gamma}}\right]}\ .
\end{equation}
Analyzing $\alpha_c$ as a function of $\tilde{\Gamma}$, we see that $\alpha_c$ vanishes when $\tilde{\Gamma}=0$ vanishes\footnote{Near $\tilde{\Gamma}=0$ we do not expect our results to be accurate since we made the assumption that $x_-\gg\sigma$ this is clearly violated as $\tilde{\Gamma}\to 0$. However we expect our results to be quite good in regimes where $\tilde{\Gamma}>1$.} and monotonically increases asymptotically as $\sim\sqrt{\tilde{\Gamma}}$ which is the same asymptotic behavior of the width of the density of states, $E_+-E_-=2\sqrt{1+\Tilde{\Gamma}}\sim \sqrt{\tilde{\Gamma}}$. So at large $N$, our computation reveals that up to a numeric pre-factor the decay of the fin is characterized by~$\sigma\sim \frac{E_+-E_-}{N}$ which we recognize as the average spacing between eigenvalues in the random sector. As we will see in later examples involving double scaled matrix models, this lesson continues to hold and the decay of the ``fin'' will also be controlled by the typical spacing between microstates in the random sector which is given by $\hbar\sim N^{-1}$.

Now that we have discussed how to fix $\sigma$ in the Gaussian approximation we can proceed and use our expressions to approximate the connected part of the spectral form factor:
\begin{equation}
    \braket{Z(\beta+it)Z(\beta-it)}_{con.}=\braket{Z_R(2\beta)}-4N^2\int_0^\infty dx_+e^{-2\beta x_+}\int_0^{x_+}dx_-\cos(2tx_-)e^{-\frac{x_-^2}{2\sigma^2}}\tilde{K}(x_+,0)^2\ .
\end{equation}
In the decoupling approximation we have:
\begin{equation}
\begin{split}
    &\braket{Z(\beta+it)Z(\beta-it)}_{con.}=\braket{Z_R(2\beta)}-4N^2\left[\int_0^\infty dx_+ e^{-2\beta x_+}\tilde{\rho}_{MP}(x_+)^2\right]\left[\int_0^\infty dx_-\cos(2tx_-)e^{-\frac{x_-^2}{2\sigma^2}}\right]\\
&=\braket{Z_R(2\beta)}-2N\alpha_c\sqrt{2\pi}\left[\int_0^\infty dx_+ e^{-2\beta x_+}\tilde{\rho}_{MP}(x_+)^2\right] e^{-\frac{2\alpha_c^2}{N^2}t^2}\\
    &=\braket{Z_R(2\beta)}-N\frac{\int_0^\infty dx_+ e^{-2\beta x_+}\tilde{\rho}_{MP}(x_+)^2}{\int_0^\infty dx_+ \tilde{\rho}_{MP}(x_+)^2}e^{-\frac{2\alpha_c^2}{N^2}t^2}\\
    &\braket{Z_R(2\beta)}=N\int_0^\infty dx_+ e^{-2\beta x_+}\tilde{\rho}_{MP}(x_+)\ .\\
\end{split}
\end{equation}
We can see that at $t=0$ at infinite temperature the form factor is exactly equal to 0 as it should in our construction. The time dependence is expressed as a Gaussian and smoothly approaches a value of $\braket{Z_R(2\beta)}$. One draw back of the Gaussian approximation to the kernel is that there is no abrupt transition to a plateau. Nonetheless, we can roughly characterize the time scale on which a plateau manifests through $\alpha_c$. In particular, we will simply define the plateau time scale to be given by twice the standard deviation of the Gaussian that appears in the form factor and write:
\begin{equation}
    t_{plateau}=\frac{N}{\alpha_c}.
\end{equation}
What we can see is that the time scale of the onset of the plateau is controlled by $N/\alpha_c$ which is characterized by the density of eigenvalues. In particular since $\alpha_c$ actually depends on $\tilde{\Gamma}$ which characterizes the number degenerate zero eigenvalues the time in which the plateau is reached is sensitive to the presence of the degenerate sector. 

\begin{figure}[h!]
    \centering
    \includegraphics[width=1\linewidth]{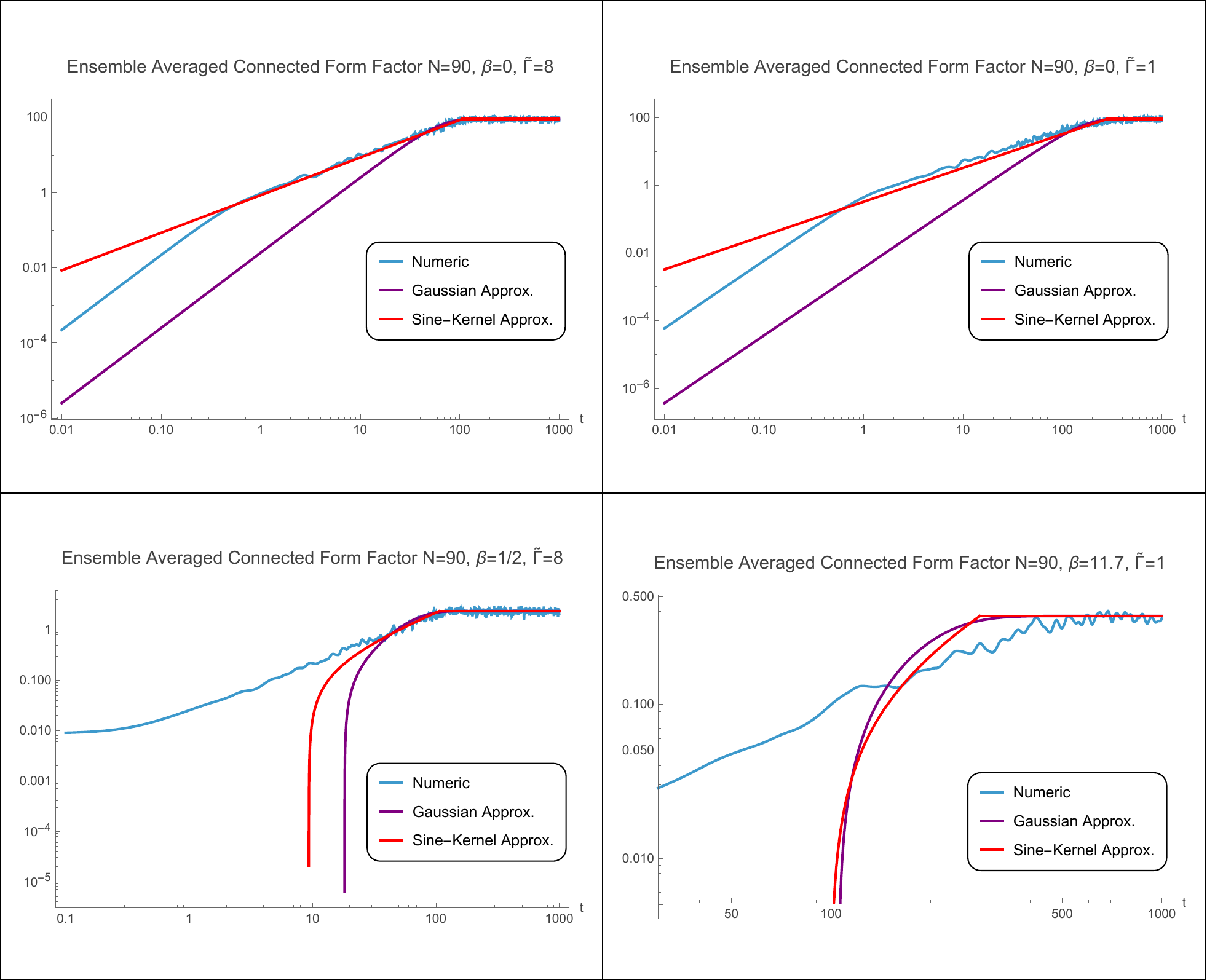}
    \caption{Comparison of computations of the spectral form factor using the Gaussian approximation (violet line) and sine-kernel approximation (red line) with numerical simulations of the ensemble averaged connected form factor (blue line) for various temperature regimes and values of $\tilde{\Gamma}$ with $N=90$ being fixed. }
    \label{TestOFApproxFigures}
\end{figure}

In Figure \ref{TestOFApproxFigures}, we compare the connected form factor obtained through this Gaussian approximation of the kernel (the violet line) to a numerical calculation of the connected form factor (blue line). What we can see is that the Gaussian approximation is capable of giving estimates of the onset time of the plateau phase however at intermediate and early times it is not a particularly accurate. Next we will consider the sine-kernel approximation.\\ 

\noindent \textbf{Sine-Kernel Approximation for Kernel:}

\noindent This type of fall off in the $x_-$ direction motivated the sine-kernel of RMT and proximate the kernel as:
\begin{equation}
    \begin{split}
        \tilde{K}(x_+,x_-)=\frac{\sin(x_-/\sigma)}{ x_-/\sigma}\tilde{K}(x_+,0)\ .
    \end{split}
\end{equation}
As before, the non-trivial task is fixing $\sigma$ according to our condition which will now read:
\begin{equation}
    \frac{1}{4N}=\int_0^\infty dx_+ \tilde{K}(x_+,0)^2\int_0^{x_+}dx_-\frac{\sin^2( x_-/\sigma)}{x_-^2/\sigma^2}\ .
\end{equation}
In the decoupling limit we have:
\begin{equation}
    \frac{1}{4N}=\frac{\pi\sigma}{2}\int_0^{\infty} dx_+ \tilde{K}(x_+,0)^2\Rightarrow \sigma=\frac{\left[\int_0^\infty dx_+ \tilde{K}(x_+,0)^2\right]^{-1}}{2\pi N}\ .
\end{equation}
In the $N\to\infty$ regime we arrive at $\sigma=\sqrt{\frac{2}{\pi}}\alpha_c N^{-1}$ where $\alpha_c$ was given in Eq. (\ref{alphaCriticalMP}). With this, we can write the following expression for the connected form factor in the decoupling approximation:
\begin{equation}
\begin{split}
    &\braket{Z(\beta+it)Z(\beta-it)}_{con.}=\braket{Z_R(2\beta)}-4N^2\int_0^\infty dx_+ e^{-2\beta x_+}\tilde{\rho}_{MP}(x_+)^2\int_0^\infty dx_-\cos(2tx_-)\frac{\sin^2(x_-/\sigma)}{x_-^2/\sigma^2}\\
    &=\braket{Z_R(2\beta)}-\frac{N}{2}\frac{\int_0^\infty dx_+ e^{-2\beta x_+}\tilde{\rho}_{MP}(x_+)^2}{\int_0^\infty dx_+ \tilde{\rho}_{MP}(x_+)^2}\left[|t\sigma-1|-(t\sigma-1)\right]\\
    &=\braket{Z_R(2\beta)}+\frac{\int_0^\infty dx_+ e^{-2\beta x_+}\tilde{\rho}_{MP}(x_+)^2}{\int_0^\infty dx_+ \tilde{\rho}_{MP}(x_+)^2}\Theta\left(\sqrt{\frac{\pi}{2}}\frac{N}{\alpha_c}-t\right)\left[\sqrt{\frac{2}{\pi}}\alpha_c t-N\right]\ .\\
\end{split}
\end{equation}
Here we see a linear ramp with slope equal to $\sqrt{\frac{2}{\pi}}\alpha_c$ accompanied by a sharp transition to a plateau which appears at $t_{plateau}=\sqrt{\frac{\pi}{2}}\frac{N}{\alpha_c}$, which is similar to what we found in the Gaussian approximation and corroborates the expectation that the plateau manifests on time scales of the inverse spacing between microstates. 

In Figure \ref{TestOFApproxFigures}, we compare how the sine-kernel approximation (red line) compares with numeric calculations of the connected part of the form factor for $N=90$ at different values of $\tilde{\Gamma}$ and $\beta$.
We see that at infinite temperature both approximations roughly capture the time of transition from a ramp to plateau. Furthermore, the sine-kernel approximation does a fairly good job at capturing the portion of the linear part of the ramp right before the plateau regime whereas the Gaussian approximation does not. At finite temperatures we still have a good estimates for the transition time to a plateau especially for larger values of $\tilde{\Gamma}$, however the departure of the approximations at sufficiently early times is far more dramatic. This dramatic behavior in the Log-Log plot at finite temperatures is because the approximated form factor actually dips below zero at sufficiently early times. Of course we know that this should not happen it should be strictly non-negative as shown by the numeric computations but our approximations cannot capture this feature. However, based on these results we can say that these relatively crude approximations work quite well especially at infinite temperatures at times leading up to the plateau phase. In particular, we have seen that the time the connected form factor will saturate to the Plateau value is sensitive to the density of eigenvalues which itself is sensitive to $\tilde{\Gamma}$ (a measure of the number of zero eigenvalues). Keeping these lessons in mind we will now move to the realm of double scaled random matrix models where more precise statements will be made.

\section{Spectral Form Factor of the Bessel Model}
\label{BesselModFFSection}
Thus far, we have considered finite $N$ random matrix ensembles. We will now analyze examples of double scaled random matrix models with gapped spectra with a large number of degenerate states. One such model, sometimes called the ``Bessel model'', \cite{Nagao1993,Carlisle:2005wa,Johnson:2020heh,Johnson:2020mwi,Johnson:2021owr,Johnson:2021rsh,Johnson:2022wsr,Johnson:2024tgg}, can be thought of as arising in a double scaled limit of the Wishart model we studied in the previous section where one zooms in on the left most edge of the spectrum \cite{Forrester1993,TracyWidom1994Bessel}. 

\subsection{Overview of the Kernel of Double Scaled Matrix Models}
Just as we did for the Wishart ensemble, our analysis will be centered around using the kernel. In particular, we will need double scaled analogues of orthogonal polynomials to construct the kernel. It turns out that, for certain classes of double scaled matrix models, which include the Bessel model as well as JT gravity and its supersymmetric extensions, the role of the orthogonal polynomials are played by wavefunctions of an auxiliary Schrodinger problem \cite{Brezin:1990rb,Douglas:1989ve,Gross:1989vs,Ginsparg:1993is,Johnson:2019eik,Johnson:2020exp,Johnson:2020heh,Johnson:2020mwi,Johnson:2021owr,Johnson:2021rsh,Johnson:2022wsr,Johnson:2023ofr,Johnson:2024bue,Johnson:2024fkm,Johnson:2024tgg,Johnson:2025vyz,Johnson:2025oty,Ahmed:2025lxe,Johnson:2026jbq}:
\begin{equation}
    \left[-\hbar^2\partial_x^2+u(x)\right]\psi(E,x)=E\psi(E,x)\ ,
\end{equation}
where the potential $u(x)$, is given by the so-called ``string equation'' given as \cite{Morris:1990cq,Dalley:1991qg,Dalley:1992br}:
\begin{equation}
    \begin{split}
         &u\mathcal{R}^2-\frac{\hbar^2}{2}\mathcal{R}\mathcal{R}''+\frac{\hbar^2}{4}\mathcal{R}'^2=\hbar^2\Gamma^2\\
        &\mathcal{R}=\sum_{k=1}^\infty t_k R_k[u(x)]+x\ ,\\
    \end{split}
\end{equation}
where $R_k[u]$ is the $k$-th Gelfand-Dikii polynomial \cite{GelfandDikii1975,GelfandDikii1977,Gelfand:1995qu}\footnote{For a more modern and explicit overview of these polynomials and their constructions the reader can look at appendices of \cite{Ahmed:2025lxe}.}. The coefficients $t_k$ are fixed to give the desired random matrix model \cite{Johnson:2019eik,Johnson:2022wsr,Johnson:2025oty}. Furthermore, the parameter $\Gamma$ in the string equation plays the role of determining the number of degenerate ground states in the model \cite{Johnson:2023ofr,Johnson:2024tgg,Johnson:2025oty,Ahmed:2025lxe}. In this formalism, the kernel is now given by the following expression \cite{Johnson:2021zuo,Johnson:2021tnl,Johnson:2022wsr,Johnson:2025dyb}:
\begin{equation}
    K(E,E')=\int \psi(E,x)\psi(E,x)dx\ ,
\end{equation}
and just as before, the diagonal of the kernel then identifies the spectral density as:
\begin{equation}
    \rho(E)=K(E,E)=\int|\psi(E,x)|^2dx\ .
\end{equation}
In particular, we will specifically be interested in the regime where $\Gamma$ scales as $\Gamma=\hbar^{-1}\tilde{\Gamma}$, this is the double scaled version of the regime we considered in the Wishart ensemble (i.e. the condition that $\Gamma=N\tilde{\Gamma}$), where the role of $N$ is now played by $\hbar^{-1}$.

\subsection{Disconnected From Factor}
The potential for the Bessel model in the formalism we reviewed in the previous section is known to be given as $u_0(x)=0+\hbar^2\left(\Gamma^2-\frac{1}{4}\right)/x^2$. As we advertised earlier, we will specifically consider a scaling where $\Gamma= \hbar^{-1}\tilde{\Gamma}$. Then, at leading order in $\hbar$ the potential reads: $u_0=\tilde{\Gamma}^2/x^2+\mathcal{O}(\hbar^2)$. Using the formula for the leading density of states expressed in terms of $u_0$ gives \cite{Johnson:2023ofr,Johnson:2024tgg}:
\begin{equation}
    \rho_0(E)=\frac{1}{2\pi\hbar} \int_{0}^{1}\frac{\Theta(E-\frac{\tilde{\Gamma}^2}{x^2})}{\sqrt{E-\frac{\tilde{\Gamma}^2}{x^2}}}dx\ ,
\end{equation}
upon evaluating the integral\footnote{It should be noted that in the wider literature (e.g. in topological recursion \cite{do2016topologicalrecursionbesselcurve,Chidambaram_2025,iwaki2018voroscoefficientshypergeometricdifferential}) the term ``Bessel model'' sometimes refers to the matrix model with a leading spectral density (or ``spectral curve'') that goes as $\rho_0=\frac{1}{2\pi\sqrt{E}}$, whereas what we refer to as the Bessel model is a matrix model which has a leading density of the form $\rho_0=\frac{1}{2\pi}\frac{\sqrt{E-E_0}}{E}$ which reduces to the other notion of ``Bessel model'' when $E_0\to 0$. We thank Clifford Johnson for pointing out this subtlety.}: 
\begin{equation}
\label{WKBDOSBessel}
    \rho_0(E)=\frac{1}{2\pi\hbar} \int_{\tilde{\Gamma}/\sqrt{E}}^{\mu=1}\frac{1}{\sqrt{E-\frac{\tilde{\Gamma}^2}{x^2}}}dx=\frac{1}{2\pi\hbar}\frac{\sqrt{E-\tilde{\Gamma}^2}}{E}\ .
\end{equation}
Explicitly including the ground states we get:
\begin{equation}
\label{ClassicalBesselDensityWithDeg}
    \hbar\rho_0(E)=\tilde{\Gamma}\delta(E) +\frac{1}{2\pi}\frac{\sqrt{E-\tilde{\Gamma}^2}}{E}\ .
\end{equation}

\begin{figure}[h!]
    \centering
    \includegraphics[width=1\linewidth]{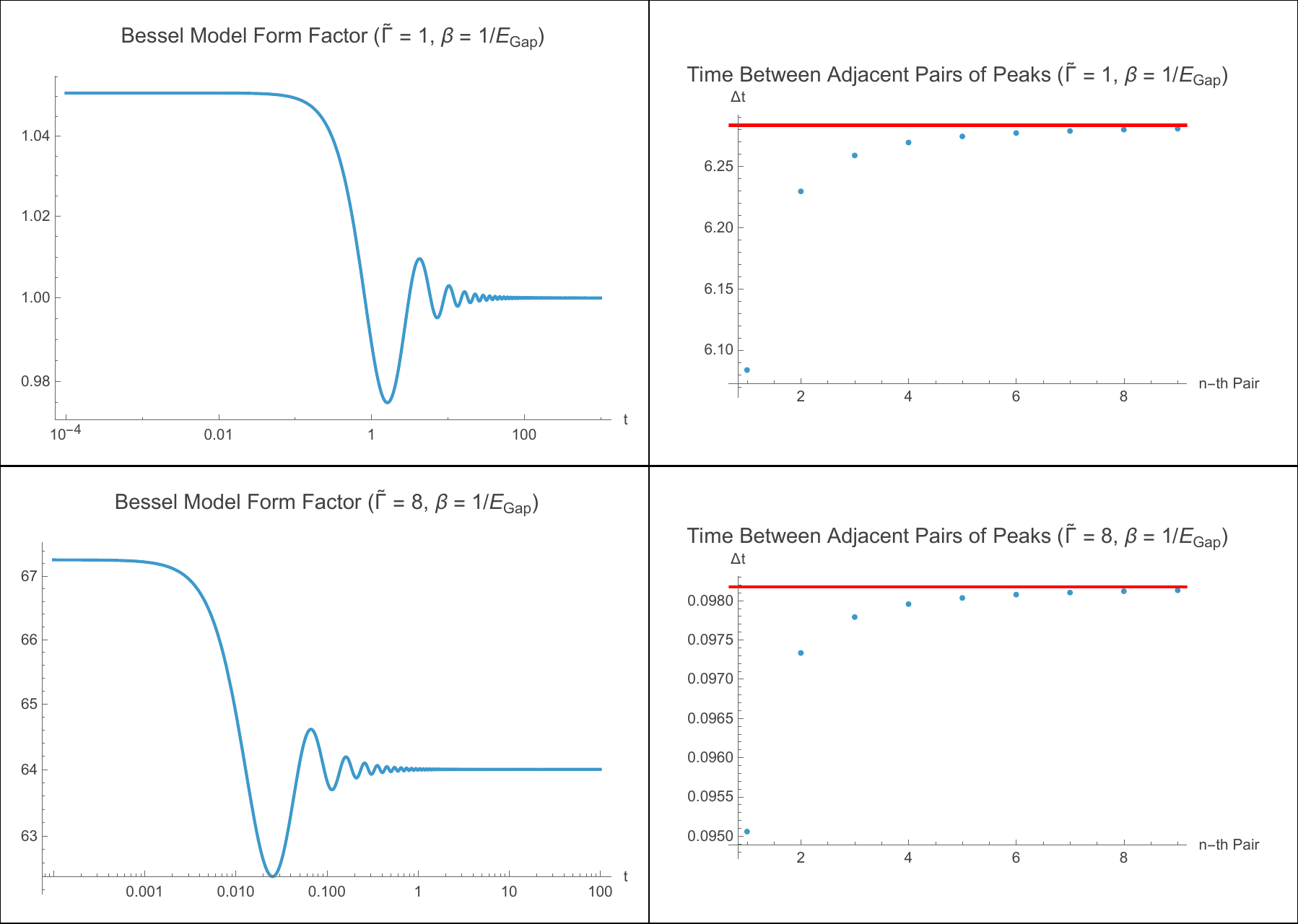}
    \caption{On the left we plot the disconnected spectral form factor for the Bessel model at temperatures equal to $E_{Gap}=\tilde{\Gamma}^2$ which reveal oscillations towards a Plateau. On the right side we do a numerical analysis of the time between adjacent peaks of the oscillations which show that they approach the discussed time scale $\tau=\frac{2\pi}{E_{Gap}}=\frac{2\pi}{\tilde{\Gamma}^2}$ at later times. Analogous to the Wishart case.}
    \label{FFPlotsBesselModel}
\end{figure}

We can clearly see that the total density of states is a combination of the degenerate sub-sector of $\Gamma$ states at $E=0$ followed by a gap of size $E_{Gap}=\tilde{\Gamma}^2$ to a continuum of random excited states\footnote{It is worth noting, that recent work \cite{Johnson:2026plw} argues that that a similar spectral density also appears as a universal model of strong BPS chaos.}. Using the leading spectral density we can study of the disconnected form factor by taking the Laplace transform of Eq. (\ref{ClassicalBesselDensityWithDeg}) to define the following partition function:
\begin{equation}
\begin{split}
    &\braket{Z(\beta)}=\frac{\tilde{\Gamma}}{\hbar}\left[1+\frac{e^{-\beta\tilde{\Gamma}^2}}{\sqrt{4\pi\beta\tilde{\Gamma}^2}}-\frac{1}{2}\text{Erfc}\left(\sqrt{\beta\tilde{\Gamma}^2}\right)\right]\ .\\
\end{split}
\end{equation}
As we can see, the expression is ill defined at $\beta=0$ so we should always consider some $\beta>0$ in our calculations. In particular, we will be interested in regimes where the temperature scale is defined by the size of the gap so that $\beta=E_{Gap}^{-1}=\tilde{\Gamma}^{-2}$. When we compute the disconnected form factor using this partition function and analyze the asymptotic expression at sufficiently large times we get an expression of the form:
\begin{equation}
    \begin{split}
        &\hbar^2\braket{Z(\tilde{\Gamma}^{-2}+it)Z(\tilde{\Gamma}^{-2}-it)}_{\text{dis.}}\sim \tilde{\Gamma}^2-e^{i\tilde{\Gamma}^2t-1}\frac{1-i}{4\tilde{\Gamma}t^{3/2}\sqrt{2\pi}}-e^{-i\tilde{\Gamma}^2 t-1}\frac{1+i}{4\tilde{\Gamma}t^{3/2}\sqrt{2\pi}}\\
        &=\tilde{\Gamma}^2-\frac{\cos\left(\tilde{\Gamma}^2 t\right)+\sin\left(\tilde{\Gamma}^2 t\right)}{2t^{3/2}\tilde{\Gamma}\sqrt{2\pi e^2}}+\mathcal{O}(1/t^{5/2})\ .\\
    \end{split}
\end{equation}
Based on this, we conclude that at sufficiently late times the form factor will exhibit decaying oscillations with a period $\tau=\frac{2\pi}{E_{Gap}}=\frac{2\pi}{\tilde{\Gamma}^2}$, which is the same result we found in the Wishart example. In Figure \ref{FFPlotsBesselModel}, we give a sample plot showing that the disconnected form factor (left plot) does indeed exhibit the expected oscillations and that the period of the oscillations approach the expected value set by the gap (right plot).  

To conclude our analysis of the disconnected part lets go back to the observation that when $\beta\to 0$ the partition function diverges. This is because in double scaled matrix models the total number of states is infinite. This fact plays an interesting role in the issue of if the connected part or disconnected part dominates the computation at late times. Based on the results in Table \ref{FFValuesWithDegens}, we know that the disconnected part of the form factor will contribute a value of $\Gamma^2$ and the connected part a value of $\braket{Z_R(2\beta)}$ at late times. From this, it is straightforward to see that at sufficiently high temperature the connected part of the form factor is the dominant contribution, in such a case we expect that the behavior of the form factor to follow the standard lore of having a dip, ramp, and plateau (i.e. the oscillations due to the gap we discussed here are highly suppressed). Nonetheless, it is interesting to ask at what temperature regimes would the late time oscillations become apparent (i.e. at what temperatures should the disconnected part of the form factor dominate over the connected part?). A straightforward way to get an estimate on this temperature is to ask when $\Gamma^2\gg \braket{Z_R(2\beta)}$. Using the classical density of states given in Eq. (\ref{ClassicalBesselDensityWithDeg}) the condition will read:
\begin{equation}
    \tilde{\Gamma}^2\gg \hbar\int_{\tilde{\Gamma}^2}^\infty \frac{\sqrt{E-\tilde{\Gamma}^2}}{2\pi E}e^{-2\beta E}dE=\frac{\hbar\tilde{\Gamma}}{2}\left[\frac{e^{-u}}{\sqrt{\pi u}}-\text{Erfc}\left[\sqrt{u}\right]\right], \qquad u=2\beta\tilde{\Gamma}^2\ .
\end{equation}
We can see due to the presence of the overall factor of $\hbar$ the temperature will typically need to be very high (when $\hbar\ll 1$) to violate this condition. We can get an estimate on this threshold temperature by expanding the right hand side of the expression to leading order in $u$ and then set it equal to $\tilde{\Gamma}^2$ and solve for $u_c=2\beta_c\tilde{\Gamma}^2$ in the equation below:
\begin{equation}
    \tilde{\Gamma}^2=\frac{\hbar\tilde{\Gamma}}{2\sqrt{\pi u_c}}\ .
\end{equation}
Isolating for $\beta_c$ and we find that $\beta_c\sim \frac{\hbar^2}{\tilde{\Gamma}^4}=\left(\frac{\hbar}{E_{Gap}}\right)^2$. Based on this, we conclude that as long as $\beta\gg\frac{\hbar^2}{E_{Gap}^2}$ then we will expect the disconnected part to dominate over the connected part. This condition is certainly true as $\hbar\to 0$ in the regimes we are interested in studying (i.e. $\tilde{\Gamma}$ is order 1 and temperature is of the order of the gap size). Therefore, we conclude that the computation of the disconnected form factor is going to essentially be the full form factor with the connected form factor being a suppressed sub-leading correction just as we saw for the Wishart ensemble. 

In the next subsection we will go beyond the leading order analysis done here and consider the full non-perturbative Bessel model to analyze the connected part of the spectral form factor and even the full non-perturbative disconnected part as well.

\subsection{Non-Perturbative Bessel Model, Kernel, and Connected Form Factor}
\begin{figure}[h!]
    \centering
    \includegraphics[width=1\linewidth]{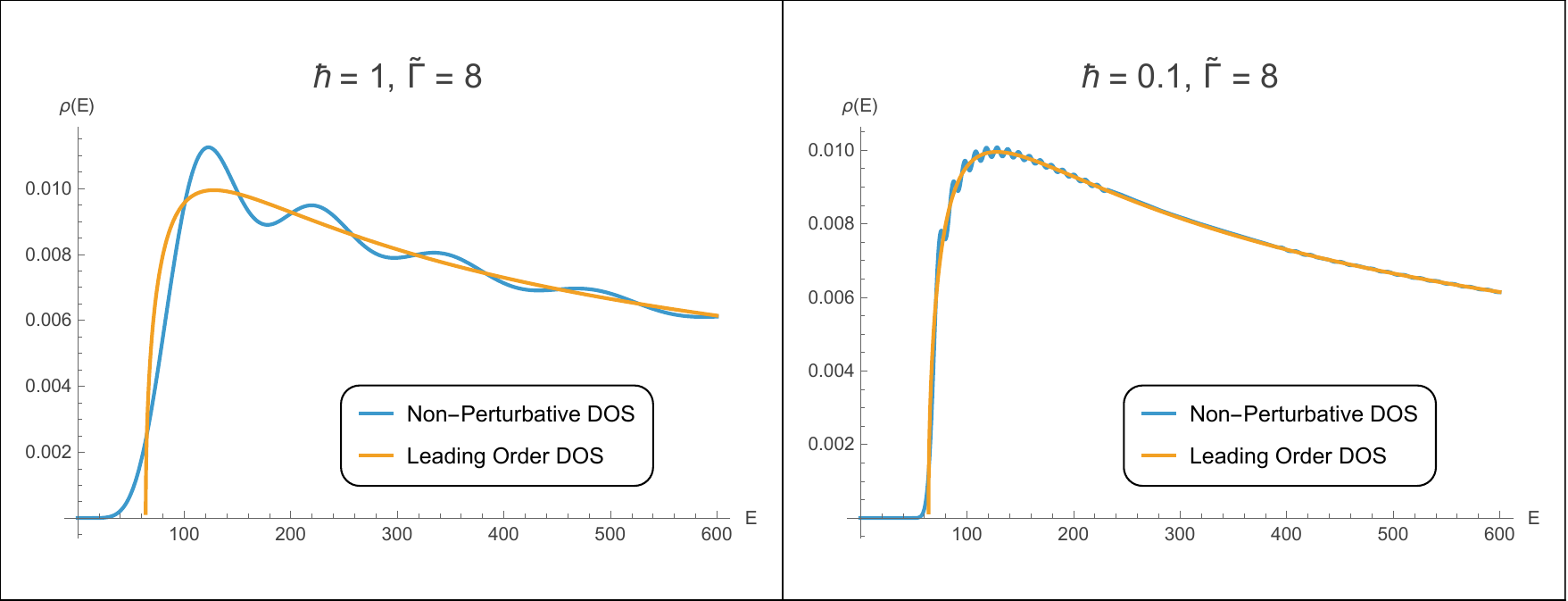}
    \caption{Here we make plots of the full non-perturbative density of states (blue line) given in Eq. (\ref{rhoBesselContinuum}) for the Bessel model for $\tilde{\Gamma}=8$ for various values of $\hbar$. We compare to the leading order density of states given by Eq. (\ref{WKBDOSBessel}). We can clearly see how as $\hbar\to 0$ we the non-perturbative density approaches the leading order density as expected. Note that these plots do not include the delta function spike of the non-random degenerate sector which should appear in the full density of states given by Eq. (\ref{rhoNonPertIncludeZeroDensity}).}
    \label{BesselNonPerVsLeadingDOS}
\end{figure}

In the case of the Bessel model we can exactly solve for the wavefunctions of the auxiliary Schrodinger problem \cite{Nagao1993,Carlisle:2005wa,Johnson:2020heh,Johnson:2020mwi,Johnson:2021owr,Johnson:2021rsh,Johnson:2022wsr,Johnson:2024tgg}:
\begin{equation}
\label{BesselSchrodinger}
    \mathcal{H}\psi=E\psi, \qquad \mathcal{H}=-\hbar^2\partial_x^2+\frac{\hbar^2\left(\Gamma^2-\frac{1}{4}\right)}{x^2} \ .
\end{equation}
The ``properly'' normalized wave functions are \footnote{The normalization is chosen such that as $\hbar\to 0$ the diagonal of the kernel constructed from these wave functions approaches the leading density of states given in Eq. (\ref{WKBDOSBessel}).}:
\begin{equation}
    \psi(E,x)=\frac{1}{\hbar}\sqrt{\frac{x}{2}}J_{\Gamma}\left(\frac{x\sqrt{E}}{\hbar}\right) \ ,
\end{equation}
where $J_{\Gamma}$ is the Bessel function of the first kind. Then the kernel is given as:
\begin{equation}
\label{BesselKernel}
\begin{split}
    &K(E,E')=\int_0^1\psi(E,x)\psi(E',x)dx\\
    &=\frac{\sqrt{E'} J_{\Gamma-1} \left(\frac{\sqrt{E'}}{\hbar}\right)J_{\Gamma} \left(\frac{\sqrt{E}}{\hbar}\right)-\sqrt{E}J_{\Gamma-1} \left(\frac{\sqrt{E}}{\hbar}\right)J_{\Gamma} \left(\frac{\sqrt{E'}}{\hbar}\right)}  {2\hbar (E-E')}\ .\\
\end{split}
\end{equation}
The non-perturbative spectral density is given by taking the limit of the kernel as $E'\to E$, this gives:
\begin{equation}
\label{rhoBesselContinuum}
    \rho(E)=\lim_{E'\to E}K(E,E')=\frac{1}{4\hbar^2}\left[J_{\Gamma}\left(\frac{\sqrt{E}}{\hbar}\right)^2-J_{\Gamma-1}\left(\frac{\sqrt{E}}{\hbar}\right)J_{\Gamma+1}\left(\frac{\sqrt{E}}{\hbar}\right)\right]\ .
\end{equation}
This only describes the continuum of states in the random sector to which we need to include the states in the degenerate sector to give the full density of states:
\begin{equation}
\label{rhoNonPertIncludeZeroDensity}
    \begin{split}
        \rho(E)=\frac{\tilde{\Gamma}\delta(E)}{\hbar}+\frac{1}{4\hbar^2}\left[J_{\tilde{\Gamma}/\hbar}\left(\sqrt{E}/\hbar\right)^2-J_{\tilde{\Gamma}/\hbar-1}\left(\sqrt{E}/\hbar\right)J_{\tilde{\Gamma}/\hbar+1}\left(\sqrt{E}/\hbar\right)\right]\ .
    \end{split}
\end{equation}
In Figure \ref{BesselNonPerVsLeadingDOS}, we compare the full non-perturbative density of states to the leading order density of states at various values of $\hbar$ and demonstrate that as $\hbar\to 0$ the non-perturbative density converges to the leading density as expected.

 Using the kernel we can revisit computations of the disconnected form factor in a fully non-perturbative regime. For example, we can compute the full non-perturbative partition function:
\begin{equation}
\label{FullPartitionFunctionBesselWithDegen}
    \begin{split}
        \braket{Z(\beta)}=\Gamma+\frac{I_\Gamma \left(\frac{1}{2\hbar^2\beta}\right)}{4\hbar^2\beta e^{\frac{1}{2\hbar^2\beta}}}-\frac{\Gamma {}_3F_3\left(\frac{1}{2}+\Gamma,1+\Gamma,1+\Gamma;\Gamma,2+\Gamma,1+2\Gamma;-\frac{1}{\hbar^2\beta}\right)}{(\Gamma+1)!\left(4\hbar^2\beta\right)^{\Gamma+1}}\ .
    \end{split}
\end{equation}
The result above can then be used to compute the disconnected form factor in a fully non-perturbative context. Furthermore, given the kernel we can also discuss the connected part of the form factor which will be the main focus of this section. 

The expression for the connected form factor is given as:
\begin{equation}
    \begin{split}
        \braket{Z(\beta+it)Z(\beta-it)}_{\text{con}.}=\braket{Z_R(2\beta)}-\int_0^\infty dE_1 \int_0^\infty dE_2 e^{-(\beta+it)E_1}e^{-(\beta-it)E_2}K(E_1,E_2)^2\ ,
    \end{split}
\end{equation}
where $K(E_1,E_2)$ is now given by the kernel in Eq. (\ref{BesselKernel}) and $\braket{Z_R(2\beta)}=\braket{Z(2\beta)}-\Gamma$ with $\braket{Z(\beta)}$ given in Eq. (\ref{FullPartitionFunctionBesselWithDegen}). Unlike in the Wishart ensemble, it is not clear that one can do the integrals exactly. Instead, we will do an analysis in the regime where $\hbar\to 0$ and try to understand if the kernel can be written in a more tractable form. As we will see the the sine-kernel will naturally emerge from such an analysis. 

\begin{figure}[h!]
    \centering
    \includegraphics[width=1\linewidth]{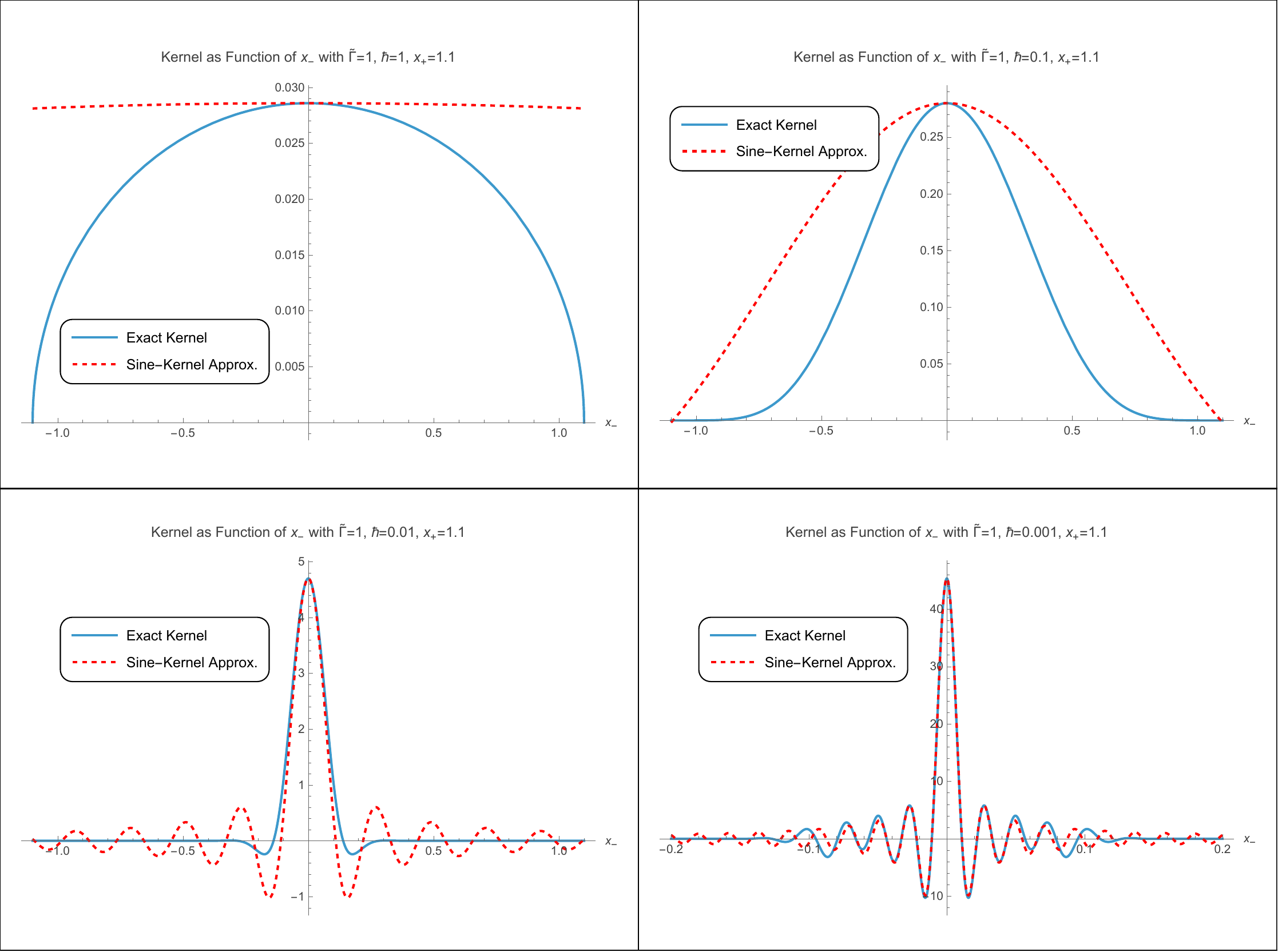}
    \caption{Here we fixed $\tilde{\Gamma}=1$ and $x_+=1.1$ and made plots of the exact Bessel model kernel (blue line) in Eq. (\ref{LightconeCoordBesselKernel}) and compare to the sine-kernel approximation (red dashed line) in Eq. (\ref{LeadingBesselSineKernel}). We can see as we lower the value of $\hbar\to 0$ the sine-kernel Approximation very accurately captures details of the exact Bessel kernel.}
    \label{ConvergenceBesselToSine}
\end{figure}

We begin by re-writing everything in a new set of variables namely the $x_{\pm}$ variables we used for the Wishart ensemble:
\begin{equation}
\label{BesselChangeOfVar}
\begin{split}
    &E_1=x_+-x_-\\
    &E_2=x_++x_-\ .\\
\end{split}
\end{equation}
In these variables, the connected form factor is given by:
\begin{equation}
\label{LightConeCoordBesselFF}
    \begin{split}
&\braket{Z(\beta+it)Z(\beta-it)}_{con.}=\braket{Z_R(2\beta)}-4\int_0^\infty dx_+ e^{-2\beta x_+} \int_0^{x_+} dx_-\cos(2tx_-)\tilde{K}(x_+,x_-)^2\ ,\\
    \end{split}
\end{equation}
with $\tilde{K}(x_+,x_-)$ given explicitly as:
\begin{equation}
\label{LightconeCoordBesselKernel}
    \begin{split}
        &\tilde{K}(x_+,x_-)=\frac{\sqrt{x_+-x_-}J_{\Gamma-1}\left(\frac{\sqrt{x_+-x_-}}{\hbar}\right)J_\Gamma\left(\frac{\sqrt{x_++x_-}}{\hbar}\right)-\sqrt{x_++x_-}J_{\Gamma-1}\left(\frac{\sqrt{x_++x_-}}{\hbar}\right)J_\Gamma\left(\frac{\sqrt{x_+-x_-}}{\hbar}\right)}{4\hbar x_-}\ .\\
    \end{split}
\end{equation}
For a fixed value of $x_+$ we can do a Taylor expansion around $x_-=0$, given as:
\begin{equation}
    \begin{split}
        \tilde{K}(x_+,x_-)=\rho(x_+)+\frac{\partial^2_{x_-}\tilde{K}(x_+,0)}{2}x_-^2+\cdot\cdot\cdot\ ,
    \end{split}
\end{equation}
where $\rho(x_+)$ is exactly given by Eq. (\ref{rhoBesselContinuum}). To understand how the kernel decays the important object to study is the second derivative of $\tilde{K}$ with respect to $x_-$ around $x_-=0$. With some work we can show that the second derivative at $x_-=0$ can be expressed in the following form:
\begin{equation}
    \begin{split}
        &\frac{\partial^2_{x_-}\tilde{K}(x_+,0)}{2}=\sum_{j=2}^4\frac{k_j(x_+)}{\hbar^j}\ ,\\
    \end{split}
\end{equation}
where $k_j(x)$ depends on $\tilde{\Gamma},\hbar,x_+$ through some products of Bessel functions. If we consider the case where $\hbar\to 0$ the dominant contribution to the second derivative should come from the term $k_4(x_+)/\hbar^4$. It turns out, that this leading term can be expressed as:
\begin{equation}
    \frac{k_4(x_+)}{\hbar^4}=-\left(\frac{x_+-\tilde{\Gamma}^2}{6\hbar^2x_+}\right)\rho(x_+)\ .
\end{equation}
So for sufficiently small $\hbar$ and near $x_-=0$  we can approximately write:
\begin{equation}
    \tilde{K}(x_+,x_-)\approx \rho(x_+)\left[1-\frac{x_+-\tilde{\Gamma}^2}{6\hbar^2 x_+^2}x_-^2+\cdot\cdot\cdot\right]\ .
\end{equation}
A particularly nice feature of the leading contribution for the small $\hbar$ limit is the appearance of the factor of $x_+-\tilde{\Gamma}^2$ which captures the sharp gap that appears in the density as $\hbar=0$. We can go even further and analyze the the dominant terms at each order in $x_-$ and to obtain a similar structure for these dominant higher order terms which can be combined and written as:
\begin{equation}
    \begin{split}
        \tilde{K}(x_+,x_-)\approx \rho(x_+)\left[1-\frac{x_+-\tilde{\Gamma}^2}{6\hbar^2 x_+^2}x_-^2+\frac{\left(x_+-\tilde{\Gamma}^2\right)^2}{120 (\hbar^2x_+^2)^2}x_-^4-\frac{(x_+-\tilde{\Gamma}^2)^3}{5040(\hbar^2x_+^2)^3}x_-^6+\cdot\cdot\cdot\right]\ .
    \end{split}
\end{equation}
Analyzing the coefficients a pattern emerges and we deduce that collecting the leading terms at each order in $x_-$ will give a series of the form:
\begin{equation}
\label{LeadingBesselSineKernel}
    \tilde{K}(x_+,x_-)\approx \rho(x_+)\sum_{k=0}^\infty \frac{(-1)^{k}}{(2k+1)!}\left(\frac{x_+-\tilde{\Gamma}^2}{\hbar^2 x_+^2}\right)^kx_-^{2k}=\rho(x_+)\frac{\sin\left(\frac{\sqrt{x_+-\tilde{\Gamma}^2}}{\hbar x_+}x_-\right)}{\frac{\sqrt{x_+-\tilde{\Gamma}^2}}{\hbar x_+}x_-}\ ,
\end{equation}
which is the sine-kernel! In fact, we recognize the expression multiplying the $x_-$ as the leading order density of states given in Eq. (\ref{WKBDOSBessel}). As an aside, one can do a similar analysis for another exactly solvable model called the Airy model and demonstrate the exact same truncated kernel also appears in that example as well\footnote{For the Airy model the potential is given by $u(x)=-x$ the leading density of states is $\rho_0(E)=\frac{1}{2\pi\hbar}\int_{-\infty}^0\frac{\Theta(E-u(x))dx}{\sqrt{E-u(x)}}=\frac{\sqrt{E}}{\pi\hbar }$. The exact wave-functions can be found and also the kernel can be expressed in term of our light-cone coordinates as $\tilde{K}(x_+,x_-)=\frac{Ai\left(\frac{x_--x_+}{\hbar^{2/3}}\right)Ai'\left(-\frac{x_-+x_+}{\hbar^{2/3}}\right)-Ai\left(-\frac{x_-+x_+}{\hbar^{2/3}}\right)Ai'\left(\frac{x_--x_+}{\hbar^{2/3}}\right)}{2x_-}$. Expanding this result around $x_-=0$ we find that at leading order the result is exactly of the form given in Eq. (\ref{BesselToSineKernel}).}. It strongly hints towards a universal result for the kernel, for classes of double scaled matrix models whose potential satisfies the sting equation, given by:
\begin{equation}
\label{BesselToSineKernel}
    \begin{split}
        \tilde{K}(x_+,x_-)\approx \rho(x_+)\frac{\sin\left(2\pi x_-\rho_0(x_+)\right)}{2\pi x_-\rho_0(x_+)}\ .
    \end{split}
\end{equation}
A subtle point to note is the appearance of both the full non-perturbative density $\rho(x_+)$ and the leading density $\rho_0(x_+)$ in this truncated sine-kernel approximation\footnote{When computing the form factor we will usually think about $\hbar$ being sufficiently small so that $\rho(x_+)\approx \rho_0(x_+)$.}. To really appreciate the emergence of the sine-kernel as $\hbar\to 0$ and how accurately it captures the full Bessel kernel we make a few sample plots in Figure \ref{ConvergenceBesselToSine}. What we clearly see is that as we lower the value of $\hbar$ the decay profile and even the oscillations away from the central peak of the full kernel is very accurately captured by the sine-kernel we derived in Eq. (\ref{BesselToSineKernel}).   

Armed with this sine-kernel we can now proceed to use it to study the leading behavior of the spectral form factor for the Bessel model. We are interested in computing:
\begin{equation}
\label{LeadingSineBesselKernelExactIntegral}
    \braket{Z(\beta+it)Z(\beta-it)}_{con.} =\braket{Z_R(2\beta)}- 4\int_0^\infty dx_+e^{-2\beta x_+}\rho(x_+)^2\int_0^{x_+} dx_- \cos(2tx_-)\frac{\sin^2(2\pi x_- \rho_0)}{4\pi^2 x_-^2 \rho_0^2}\ .
\end{equation}
This simplifies in the decoupling approximation, where we take the upper bound in the integral with respect to $x_-$ to infinity and write:
\begin{equation}
    \begin{split}
        &\braket{Z(\beta+it)Z(\beta-it)}_{con.}=\braket{Z_R(2\beta)}-4\int_0^\infty dx_+e^{-2\beta x_+}\rho(x_+)^2\int_0^\infty dx_- \cos(2tx_-)\frac{\sin^2(2\pi x_- \rho_0)}{4\pi^2 x_-^2 \rho_0^2}\\
        &=\braket{Z_R(2\beta)}-4\int_0^\infty dx_+e^{-2\beta x_+} \rho(x_+)^2\left[\frac{-t+2\pi \rho_0 +|t-2\pi\rho_0|}{16\pi \rho_0^2}\right]\\
        &=\braket{Z_R(2\beta)}+\frac{1}{2\pi}\int_0^\infty dx_+ e^{-2\beta x_+}\left[t-2\pi\rho_0(x_+)\right]\Theta\left(2\pi \rho_0(x_+)-t\right)\ , \\
    \end{split}
\end{equation}
where in going from the second to final line we approximated $\rho(x_+)^2/\rho_0(x_+)^2\approx 1$, which we expect to be valid when $\hbar$ is sufficiently small. Due to the presence of the step function we should view this integral as effectively having time dependent upper and lower bounds. At $t=0$ and $\beta>0$, we can see that the integral will exactly evaluate to $-\braket{Z_R(2\beta)}$ which implies that in the decoupling approximation the connected form factor exactly vanishes\footnote{Of course this is an artifact of the decoupling approximation we used, in reality we expect the form factor at $t=0$ to be some positive value.}. On the other hand, we know that $\rho_0(x_+)$ will have a local maximum. We can exactly compute its location at $x_+=2\tilde{\Gamma}^2$ and we find $2\pi\rho_0(x_+)|_{x_+=2\tilde{\Gamma}^2}=(2\hbar\tilde{\Gamma})^{-1}$. This implies that when $t=t_{plateau}=(2\hbar\tilde{\Gamma})^{-1}$ a sharp transition into the plateau plateau phase will occur, which explicitly depends on the number of degenerate states measured by $\tilde{\Gamma}$. At intermediate times, $t\in(0,t_{plateau})$, we need to explicitly solve $2\pi\rho_0(x_+)-t=0$. This can also be done exactly and there are two solutions:
\begin{equation}
    \epsilon_\pm(t)=\frac{1\pm\sqrt{1-4\hbar^2\tilde{\Gamma}^2t^2}}{2\hbar^2t^2}\ .
\end{equation}
So the for $t\in[0,t_{plateau}]$ we have the following expression for the connected form factor:
\begin{equation}
\label{DecoupledApproxBesselConFF}
\begin{split}
    &\braket{Z(\beta+it)Z(\beta-it)}_{con.}=\braket{Z_R(2\beta)}+\frac{1}{2\pi}\int_{\epsilon_-}^{\epsilon_+}dx_+e^{-2\beta x_+} \left[t-2\pi \rho_0(x_+)\right]\\
    &=\braket{Z_R(2\beta)}+\frac{e^{-2\beta\epsilon_-}-e^{-2\beta \epsilon_+}}{4\pi\beta}t-\int_{\epsilon_-}^{\epsilon^{+}}dx_+ e^{-2\beta x_+}\rho_0(x_+)\\
    &=\int_{\tilde{\Gamma^2}}^{\epsilon_-}dx_+ e^{-2\beta x_+}\rho_0(x_+)+\int_{\epsilon^+}^\infty dx_+ e^{-2\beta x_+}\rho_0(x_+)+\frac{e^{-2\beta\epsilon_-}-e^{-2\beta \epsilon_+}}{4\pi\beta}t\ . \\
\end{split}
\end{equation}
We can see one integral describes things near the gap and the other integral describes aspects far away. Also, note every single term has non-trivial time dependence due to the presence of $\epsilon_{\pm}$ which are time dependent. Now lets consider what happens at early times near $t=0$. In this case we effectively ignore the integrals and the expand the the remaining term at $t=0$ (in this expansion at leading order $\epsilon_-=\tilde{\Gamma}^2$ and $\epsilon_+\to\infty $) to obtain the following linear behavior for the ramp:
\begin{equation}
    \braket{Z(\beta+it)Z(\beta-it)}_{con.}=\frac{e^{-2\beta \tilde{\Gamma}^2}}{4\pi\beta} t\ .
\end{equation}
In particular, we can can clearly see how the slope of the connected form factor of the ramp depends on $\tilde{\Gamma}^2$.\footnote{The appearance of this result is particularly encouraging due to the fact that $\tilde{\Gamma}^2$ is exactly the size of the gap. This result exactly matches what one would obtain by computing the slope of the ramp using the leading double trumpet which we will discuss in more detail in the context of $\mathcal{N}=2$ JT supergravity.} In the regime where $\beta= 1/\tilde{\Gamma}^2$ we have:
\begin{equation}
    \braket{Z(\beta+it)Z(\beta-it)}_{con.}=\frac{\tilde{\Gamma}^{2}}{4\pi e^2}t\ .
\end{equation}
Based on this we can roughly estimate the height of the plateau in this regime to be of the order $\mathcal{O}\left(\tilde{\Gamma}/\hbar\right)$ which when compared to the late time value of $\tilde{\Gamma}^2/\hbar^2$ of the disconnected part is sub-leading which verifies our claim that when $\beta=\tilde{\Gamma}^{-2}$ the form factor is dominated by the disconnected contribution. 

\begin{figure}[h!]
    \centering
    \includegraphics[width=1\linewidth]{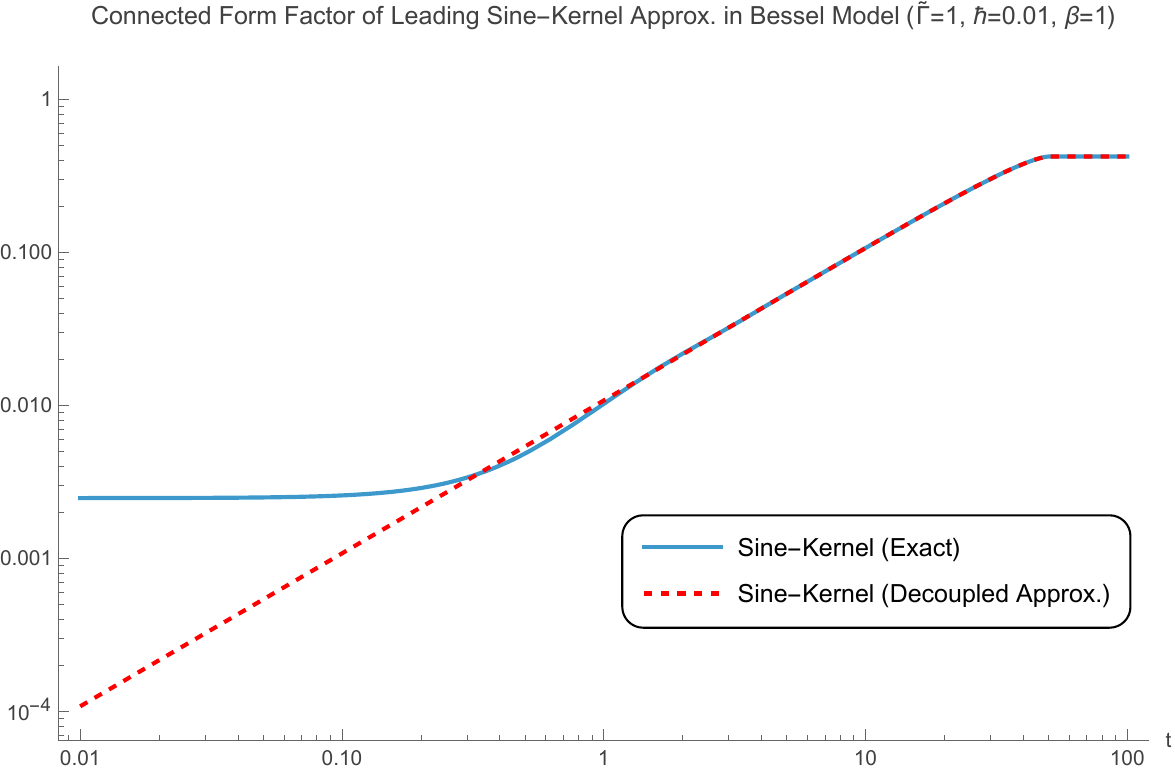}
    \caption{We plot the connected spectral form factor associated to the sine-kernel given in Eq. (\ref{BesselToSineKernel}). The exact evaluation of the expression in Eq. (\ref{LeadingSineBesselKernelExactIntegral}) gives the blue line and the decoupled approximation in Eq. (\ref{DecoupledApproxBesselConFF}) gives the dashed red line. The slope of the linear part of the dashed red line to leading order is exactly $\frac{e^{-2\beta \tilde{\Gamma}^2}}{4\pi\beta}$. In this plot we use $\tilde{\Gamma}=1,\hbar=0.01,\beta=1$.}
    \label{DecoupVdNonDecoupCalc}
\end{figure}

Now that we have discussed the leading contribution in the $\hbar\to 0$ regime in the so-called decoupling limit we will also ask what the result should be if we did not use the decoupling limit (i.e. we actually do the integral in Eq. (\ref{LeadingSineBesselKernelExactIntegral})). This can be done numerically. As an example, we compare the exact computation of the expression involving the sine-kernel with the decoupled approximation we just used for $\hbar=0.01, \tilde{\Gamma}=1, \beta=1$ in Figure \ref{DecoupVdNonDecoupCalc}.   

What we can see is that the decoupled approximation is not particularly great at capturing the early time behavior of the form factor. However, it is very good at describing the result at intermediate times where the ramp description becomes relevant all the way to the plateau. These results echo the conclusions we reached in our analysis of the Wishart ensemble using the fitted sine-kernel, in the sense that, the decoupling approximation will generally do a poor job towards approximating the form factor at sufficiently early times but is quite adept at capturing aspects about the ramp and its transition to a plateau.

\section{Spectral Form Factor of $\mathcal{N}=2$ JT Supergravity}
\label{N=2Section}
In this section, we will analyze another double scaled matrix model which corresponds to $\mathcal{N}=2$ JT supergravity. In this case the potential in the Schrodinger formalism for this theory is described by solutions to the following string equation \cite{Johnson:2023ofr,Johnson:2024tgg,Johnson:2025oty,Ahmed:2025lxe}:
\begin{equation}
\label{N=2SJTStringEq}
    \begin{split}
        &u\mathcal{R}^2-\frac{\hbar^2}{2}\mathcal{R}\mathcal{R}''+\frac{\hbar^2}{4}\mathcal{R}'^2=\tilde{\Gamma}^2\\
        &\mathcal{R}=\sum_{k=1}^\infty t_k R_k[u(x)]+x\\
        &t_k=\frac{\pi^{k-1}J_k(2\pi\sqrt{E_0})}{2(2k+1)k!E_0^{k/2}}\\
        &\mu=t_0=\frac{J_0(2\pi\sqrt{E_0})}{2\pi}\ .\\
    \end{split}
\end{equation}
As one can clearly see there the solution to the string equation which needs to be fed into the Schrodinger equation potential will be far more complicated than the simple potential of the Bessel model, this represents a major obstacle towards obtaining exact results for the wavefunction to construct the kernel. Nonetheless, we know that at leading order, the disk spectral density is reproduced. This on its own is enough to compute the disconnected part of the spectral form factor (at least in regimes where $\hbar=e^{-S_0}\to 0$). As for the analysis of the connected part, we will provide a conjectured expression for the form factor inspired by our analysis in the Bessel model. We will further demonstrate how the conjecture reproduces the exact slope of the ramp (as predicted by gluing two $\mathcal{N}=2$ trumpets together) and also provides information on how the ramp  will transition to a plateau phase which is not accessible to any finite order in the genus expansion.

\subsection{Disconnected Form Factor}
\begin{figure}[h!]
    \centering
    \includegraphics[width=1\linewidth]{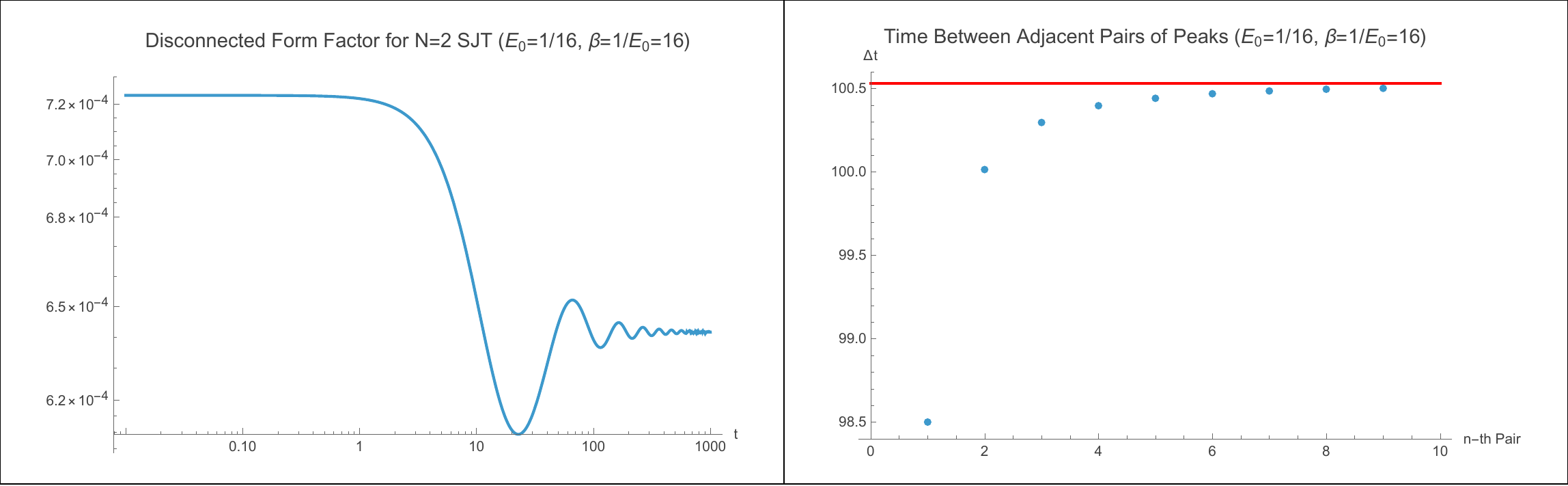}
    \caption{On the left we give a sample plot of the disconnected form factor using the leading leading density of states explicitly including the zero states. On the right we give the time difference between the $n$-th pair of adjacent peaks given by the blue points. We can see that they asymptotically approach the red line which is exactly given by $\tau=\frac{2\pi}{E_0}$, demonstrating that the oscillations in the form factor approach a period controlled by the gap size. This particular, plot was made by fixing the parameters $E_0=\frac{1}{16} \Rightarrow \tilde{\Gamma}=\frac{1}{4\pi^2}$ and $\beta=\frac{1}{E_0}=16$.}
    \label{N=2FFAndOscillAnalysisPlot}
\end{figure}

In preparation towards studying the disconnected form factor we give a brief overview of the leading (disk) density of states for $\mathcal{N}=2$ JT supergravity for a fixed $R$-charge sector \cite{Turiaci:2023jfa,Johnson:2023ofr,Johnson:2024tgg,Johnson:2025oty,Ahmed:2025lxe}:
\begin{equation}
\begin{split}
    &\hbar\rho_0(E)=\tilde{\Gamma}\delta(E)\Theta(1-4E_0)+\frac{1}{2\pi}\frac{\sinh\left(2\pi\sqrt{E-E_0}\right)}{4\pi^2 E}\Theta(E-E_0)\\
    &\tilde{\Gamma}=\frac{\sin(2\pi\sqrt{E_0})}{4\pi^2}\ .\\
\end{split}
\end{equation}
Here we see the relation between the number of BPS states (i.e. $\Gamma=\hbar^{-1}\tilde{\Gamma}$) and the size of the gap, $E_0$, is not monotonic like it was in the Wishart and Bessel models. Furthermore, we see that BPS states only exist when $0<E_0<1/4$. Based on these observations we can write the partition function associated to the density of states as:
\begin{equation}
    \braket{Z(\beta)}=\frac{\tilde{\Gamma}}{\hbar}\Theta(1-4E_0)+\int_{E_0}^\infty \frac{1}{2\pi\hbar}\frac{\sinh(2\pi\sqrt{E-E_0})}{4\pi^2 E}e^{-\beta E}dE\ .
\end{equation}
Since the integral cannot be done analytically we will resort to numerical computations to analyze the disconnected form factor. In particular, we will be interested in numerical computations in the specific regime where the temperature we consider is of the order of the size of the gap (i.e. $\beta=1/E_0$). Unsurprisingly at this point, we again find the existence of oscillations whose period asymptotically approaches $\tau=\frac{2\pi}{E_0}$. In Figure \ref{N=2FFAndOscillAnalysisPlot}, we give sample plots which illustrate these findings. Furthermore, we also expect that at sufficiently low temperatures of the order of the size of the gap, the disconnected part of the form factor will dominate the form factor computation. 

\subsection{Connected Form Factor, Wormholes, and the Kernel}
To discuss our analysis of the connected part of the spectral form factor it is useful to recall a number of well known facts. Namely that, the connected part of the form factor at leading order is captured by the so called ``wormhole'' contribution. This is given by gluing together two trumpets \cite{Banks:1989df,Ginsparg:1993is,Turiaci:2023jfa,Ahmed:2025lxe}:
\begin{equation}
\begin{split}
    &\braket{Z(\beta_1)Z(\beta_2)}_{\text{con.}}=\int_0^\infty bdbZ_{\text{Tr}}(\beta_1,b)Z_{\text{Tr}}(\beta_2,b)\\
    &Z_{\text{Tr}}(\beta,b)=\frac{1}{\sqrt{4\pi\beta}}e^{-\frac{b^2}{4\beta}}e^{-\beta E_0}\ .\\
\end{split}
\end{equation}
After performing the integral and analytically continuing to $\beta_1=\beta+it$ and $\beta_2=\beta-it$ we obtain the the connected spectral form factor at leading order in the genus expansion:
\begin{equation}
\label{SJTGravConnFF}
    \braket{Z(\beta+it)Z(\beta-it)}_{\text{con.}}=\frac{e^{-2\beta E_0}}{4\pi\beta}\sqrt{\beta^2+t^2}\ .
\end{equation}
This recovers the famous linear ramp of the form factor that characterizes eigenvalue repulsion. A critical observation of this leading order result is that it appears to continue to grow without bounds. Of course, this is not truly the case and it is believed that fully accounting for higher order genus effects and/or non-perturbative effects will transition the ramp to a plateau at sufficiently late times \cite{Saad:2019lba,Johnson:2020exp,Okuyama:2020ncd,Saad:2022kfe,Blommaert:2022lbh,Griguolo:2023jyy,Okuyama:2023pio}. Schematically, given the two geodesic boundary Weil-Petersson volumes of $\mathcal{N}=2$ SJT, which we will denote as~$V_{g,2}(b_1,b_2)$, we should compute the following series:
\begin{equation}
\begin{split}
    &\braket{Z(\beta+it)Z(\beta-it)}_{con.}=\frac{e^{-2\beta E_0}}{4\pi\beta}\sqrt{\beta^2+t^2}\\
    &+\hbar^{2g}\sum_{g=1}^\infty\int_0^\infty db_1\int_0^\infty db_2 Z_{Tr}(\beta+it,b_1)V_{g,2}(b_1,b_2)Z_{Tr}(\beta-it,b_2)\\
    &+\text{non-perturb.}\ .\\
\end{split}
\end{equation}
From this, the ramp is captured by the zeroth order term in the genus expansion and resuming all the higher order genus effects and as well as non-perturbative terms should describe the transition of the ramp to a plateau. Generally, resuming such a series is currently out of reach but it has been argued that the result should be viewed as an asymptotic series which formally fails to converge (this would likely be remedied by fully understanding the non-perturbative sector beyond the genus expansion) \cite{Kimura:2020zke,Eynard:2023qdr,Johnson:2026jbq}. It is beyond the scope and not the focus of this current work to fully understand such subtitles. However, there are still interesting comments and conjectures we can make on the questions regarding the transition from the ramp to plateau based on our computations thus far, in the context of the Schrodinger formalism for this theory. 

\begin{figure}[h!]
    \centering
    \includegraphics[width=1\linewidth]{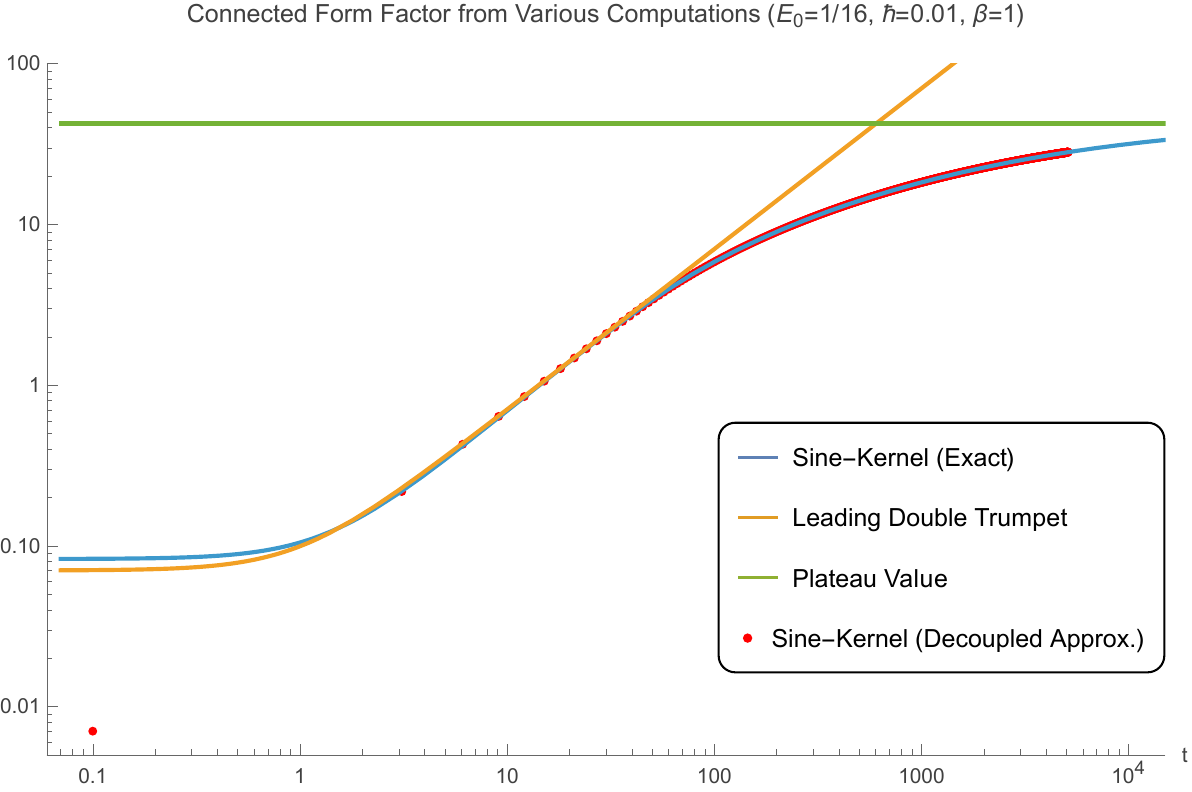}
    \caption{We plot different computations of the connected form factor for $\mathcal{N}=2$ SJT gravity. The blue line is what we get by numerically evaluating the conjectured sine-kernel approximation in Eq. (\ref{ExactSineKernelIntegralN=2SJT}). We compare this to the decoupled approximation this was done by evaluating Eq. (\ref{SineKerDecoupledN=2SJT}) for a discrete set of times and the results are given by the red points on the plot. As expected we see the exact numerical evaluation of the sine-kernel form factor to agree with the decoupled result along the ramp phase all the way to the plateau whose height is given by the green line. Finally we compared these sine-kernel approximations of the form factor to the leading order double trumpet result given by the orange line. The fixed parameters for this plot are $E_0=1/16,\hbar=0.01,\beta=1$.}
    \label{N=2ConnFFConject}
\end{figure}

To motivate our discussions, let us return to the analysis of the Bessel model which can be exactly solved non-perturbatively (via the exact solutions of the auxiliary Schrodinger problem $\psi(E,x)$ where $x\in(0,1)$ and $E>0$). As we already mentioned at the beginning of this section, the same Schrodinger formalism can also incorporate $\mathcal{N}=2$ JT supergravity and in fact provides a complete non-perturbative description beyond the genus expansion. This opens up the possibility of understanding the transition from the ramp to plateau in such a framework. However, a major hurdle towards this goal is our lack of understanding of the exact form of the wavefunctions involved. Nonetheless, we can still formally write expressions for the kernel in terms of the exact wave functions. We will denote them as,~$\psi_{SJT}(E,x)$. They will be taken to be the solutions to the Schrodinger problem with a potential that solves Eq. (\ref{N=2SJTStringEq}). Given this, construction of the kernel will be analogous to other cases we have studied:
\begin{equation}
    K_{SJT}(E_1,E_2)=\int_{-\infty}^{\mu=t_0} dx \psi_{SJT}(E_1,x)\psi_{SJT}(E_2,x)\ .
\end{equation}
So far in our setup, is the same as the Bessel case except the wavefunctions are modified. Next consider doing the same change of variables as in the Bessel case given by Eq. (\ref{BesselChangeOfVar}). Then, when we do an expansion of kernel around $x_-=0$ we would expect to get a series expansion with a structure similar to that of the Bessel model. More specifically, we conjecture that at each order in the expansion $x_-$ the leading terms for sufficiently small $\hbar$ can be resummed to give a sine-kernel of the form:
\begin{equation}
    \begin{split}
        &K_{SJT}(x_+,x_-)\approx \frac{\sin\left(2\pi x_-\rho_{0,SJT}(x_+)\right)}{2\pi x_-}\\
        &\rho_{0,SJT}(x_+)=\frac{1}{2\pi\hbar}\frac{\sinh\left(2\pi\sqrt{x_+-E_0}\right)}{4\pi^2x_+}\ .\\
    \end{split}
\end{equation}
Using this conjectured form of the kernel for $\mathcal{N}=2$ SJT the expression for the connected form factor will read:
\begin{equation}
\label{ExactSineKernelIntegralN=2SJT}
    \braket{Z(\beta+it)Z(\beta-it)}_{con.}=\braket{Z_R(2\beta)}-4\int_0^\infty dx_+ e^{-2\beta x_+}\int_0^{x_+} dx_-\cos(2tx_-)\frac{\sin^2\left(2\pi x_- \rho_{0,SJT}\right)}{4\pi^2 x_-^2}\ .
\end{equation}
In the decoupled approximation we have:
\begin{equation}
    \begin{split}
        &\braket{Z(\beta+it)Z(\beta-it)}_{con.}\\
        &=\braket{Z_R(2\beta)}+\frac{1}{2\pi}\int_0^\infty dx_+e^{-2\beta x_+}\left[t-2\pi\rho_{0,SJT}(x_+)\right]\Theta\left(2\pi \rho_{0,SJT}(x_+)-t\right)\ .\\
    \end{split}
\end{equation}
A major difference between the Bessel and $\mathcal{N}=2$ JT supergravity density of states is the lack of a global maximum at finite $x_+$ for the latter example. Due to this, there will be no sharp transition to a plateau from the ramp as we saw in the Bessel model\footnote{This same effect, where sharp transitions from ramp to plateau occur when the spectral density is bounded, was also discussed in \cite{Forrester_2021}.}. The next major difference is the number of solutions to $2\pi \rho_{0,SJT}-t=0$ there will be one such solution which we will denote as $x_*(t)$ and will play the role of the lower bound on the integral defining the connected form factor. Using it we can write: 
\begin{equation}
\label{SineKerDecoupledN=2SJT}
\begin{split}
    &\braket{Z(\beta+it)Z(\beta-it)}_{con.}=\braket{Z_R(2\beta)}+\frac{1}{2\pi}\int_{x_*}^\infty dx_+ e^{-2\beta x_+}\left[t-2\pi\rho_{0,SJT}(x_+)\right]\\
    &=\braket{Z_R(2\beta)}-\int_{x_*}^\infty dx_+e^{-2\beta x_+}\rho_{0,SJT}(x_+)+\frac{e^{-2\beta x_*(t)}}{4\pi\beta}t\\
    &=\frac{e^{-2\beta x_*(t)}}{4\pi\beta}t+\int_{E_0}^{x_*(t)} dx_+ e^{-2\beta x_+}\rho_{0,SJT}(x_+)\ .\\
\end{split}  
\end{equation}
Next we need to address the expression for $x_*(t)$. It will be given by solving:
\begin{equation}
    \frac{\sinh\left(2\pi\sqrt{x_*-E_0}\right)}{ x_*}=4\pi^2\hbar t\ .
\end{equation}
This is a transcendental equation for $x_*$ and cannot be solved exactly. However, we can see the trivial solution is at $t=0$ where $x_*(0)=E_0$. Using this, we recover the linear ramp with slope $\frac{e^{-2\beta E_0}}{4\pi\beta}$ which is consistent with the leading double trumpet result. Moreover, we can also see how the slope depends on $\tilde{\Gamma}$ since there is a well defined relation between $E_0$ and $\tilde{\Gamma}$. For more general times the computations become more difficult to do analytically so we will resort to numerical computations. This will involve fixing a specific value of $t$ then solving the transcendental equation numerically and then substituting the value of $x_*$ as a lower bound in to the integral which is then itself numerically integrated. We can also do numerical computations of the exact integral without using the decoupling approximation as well. In Figure \ref{N=2ConnFFConject}, we give a sample plot for the connected spectral form factor as given by the different approaches we used to calculate it. 

Similar to the computations for the Bessel model, the exact sine-kernel (blue line) and decoupled approximation (red dots) are in close agreement at intermediate and late times describing the linear ramp and its transition to a plateau. We can also see that at time scales where the form factor exhibits a linear ramp there is very close agreement with the leading double trumpet result (orange line) reviewed in Eq. (\ref{SJTGravConnFF}).

A peculiar feature about Figure \ref{N=2ConnFFConject} is the mismatch between the leading trumpet result and the sine-kernel result at $t=0$. Naively, since the sine-kernel expression we used is conjectured to be the leading order result in the $\hbar\to 0$ computation we should expect agreement for sufficiently small $\hbar$. We checked this and found that as we consider smaller values of $\hbar$ the result does not converge towards the double trumpet computation at $t=0$. This begs the question of what may have gone wrong and why the sine-kernel truncation is adept at capturing the ramp phase but not the early time phase? To explore this question it is helpful to try to derive expressions for the $t=0$ value of the conjectured form of the expression for the connected form factor given in Eq. (\ref{ExactSineKernelIntegralN=2SJT}). When we set $t=0$ we obtain the following expression:
\begin{equation}
    \lim_{t\to 0}\braket{Z(\beta+it)Z(\beta-it)}_{con.}=\braket{Z_R(2\beta)}-4\int_0^\infty dx_+ e^{-2\beta x_+}\int_0^{x_+} dx_-\frac{\sin^2\left(2\pi x_- \rho_{0,SJT}\right)}{4\pi^2 x_-^2}\ .
\end{equation}
We can evaluate the integral in $x_-$ in terms of special functions to obtain:
\begin{equation}
\begin{split}
    &\lim_{t\to 0}\braket{Z(\beta+it)Z(\beta-it)}_{con.}\\
    &=\braket{Z_R(2\beta)}-4\int_{E_0}^\infty dx_+ e^{-2\beta x_+}\left[\frac{-1+\cos(4\pi x_+\rho_0)}{8\pi^2 x_+}+\frac{\rho_0}{2\pi} \text{Si}\left(4\pi x_+ \rho_0\right)\right]\\
    &=\int_{E_0}^\infty dx_+e^{-2\beta x_+}\left[\rho_0-\frac{2\rho_0}{\pi}\text{Si}(4\pi x_+ \rho_0)+\frac{1-\cos(4\pi x_+\rho_0)}{2\pi^2 x_+}\right]\ ,\\
\end{split}
\end{equation}
where in the last line we combined the integral of $\braket{Z_R(2\beta)}$ with the remaining integral. In the limit where $\hbar\to 0$ we have $\rho_0\to\infty$ and the integrand simplifies a because~$\lim_{x\to\infty}\text{Si}(x)\to\pi/2$ resulting in a cancellation of the first two terms in the integrand. For the final term involving $1-\cos(4\pi x_+\rho_0)$, we effectively average over the very fast oscillations to give zero for the $\cos$ and all that remains is an integral of the form:
\begin{equation}
   \lim_{t\to 0}\braket{Z(\beta+it)Z(\beta-it)}_{con.}\approx  \int_{E_0}^\infty dx_+ \frac{e^{-2\beta x_+}}{2\pi^2 x_+}=\frac{\Gamma[0,2\beta E_0]}{2\pi^2}\ ,
\end{equation}
where $\Gamma[0,2\beta E_0]$ is the upper incomplete Gamma function. One can check that this approximation very accurately captures the value of the form factor as approximated by the sine-kernel for sufficiently small $\hbar$ at $t=0$. An interesting observation to consider is the asymptotic expansion as $\beta E_0\to \infty$ the leading contribution is\footnote{This leading approximation does quite surprising well in approximating the full function even in regimes where $\beta E_0\geq 1$.}:
\begin{equation}
    \frac{\Gamma[0,2\beta E_0]}{2\pi^2}\sim \frac{1}{\pi\beta E_0}\left(\frac{e^{-2\beta E_0}}{4\pi}\right)\ ,
\end{equation}
here we wrote it so that that it is clear the term in brackets is the exactly the leading double trumpet result at $t=0$. What we see is the sine-kernel gives a different result up to a pre-factor roughly characterized by $1/(\pi\beta E_0)$. Now that we have a some understanding of how the sine-kernel approximation differs from the double trumpet result at $t=0$ we will discuss what might correct the result. Start by noting that the sine-kernel is given by throwing away all the sub-leading terms in the $\hbar\to 0$ regime in the expansion of the kernel around $x_-=0$. Perhaps these sub-leading non-perturbative terms become important at very early times but are less important at later times. This would explain why the approximation does a good job at later times where the ramp description is relevant and one may speculate it would still be good when describing the transition to the plateau up to some small corrections (in the $\hbar\to 0$ regime) that are difficult to precisely quantify given what we know. 

To conclude the discussions of this section, it is worth noting that the derivations and results discussing the connected form factor of the double scaled matrix models we have discussed thus far (i.e. Bessel, Airy, $\mathcal{N}=2$ JT supergravity) appear to naturally embed in recent discussions of the so called $\tau$-scaling limit of the spectral form factor \cite{Okuyama:2020ncd,Saad:2022kfe,Blommaert:2022lbh,Okuyama:2023pio}. The central idea behind these works is that certain universal aspects of the transition between the ramp and plateau can be studied by taking $t\to\infty$ and $\hbar\to 0$ while holding the parameter $\tau=\hbar t$ fixed. It is argued that in this regime the spectral form factor is described by universal short range eigenvalue correlations given by the sine-kernel and it is possible to study corrections to this leading result and interpret them geometrically in a perturbative genus expansion. Therefore, we can see that the sine-kernel truncation we obtained explicitly in the Bessel and Airy models and then conjectured for $\mathcal{N}=2$ JT supergravity may be reinterpreted as the leading order result in \cite{Okuyama:2023pio}, and gives another complementary path to understand the emergence of the sine-kernel as $\hbar\to 0$. Furthermore, since the $\tau$-scaled form factor is a perturbation theory which is actually centered around~$t\to\infty$ we can see that it may not be able to completely capture the early time behavior near $t=0$. We will make some additional comments on this in the conclusion of this work.

\section{Conclusions and Future Directions}
\label{ConcludeSec}
In this work, we studied the form factor of random matrix systems with a parametrically large number of degenerate ground states which generate macroscopic gaps in the spectra of states. Through our studies, we uncovered a number of interesting results which modify the standard story of the spectral form factor. 

The results of Section \ref{GenAspFFSection} highlighted two important results. The first showed that the disconnected form factor cannot decay all the way to zero at arbitrarily late times and will generally settle to a value given by $\Gamma^2$ (i.e. the square of the number of degenerate states). The second result demonstrated that even in the presence of degenerate sub-sectors, the connected part of the form factor only depended the statistics of the eigenvalues in the non-degenerate sector. Together these two results implied that at arbitrarily late times, if $\Gamma$ scaled with the number of non-degenerate eigenvalues (or if the temperature was sufficiently low), the disconnected contribution to the form factor was the dominant contribution when compared to the connected part (a dramatic departure from the standard lore of the spectral form factor). 

In the proceeding Sections \ref{SFFWishartSection}-\ref{N=2Section} we studied explicit examples (Wishart, Bessel, and $\mathcal{N}=2$ JT supergravity models) which corroborated the findings of Section \ref{GenAspFFSection}. Through the course of studying the form factor in these examples we additionally uncovered the existence of damped oscillations whose period was generally given by $\tau=\frac{2\pi}{E_{Gap}}$, where $E_{Gap}$ is the size of the macroscopic gap which generally depends on $\Gamma$ in model specific ways. Furthermore, we also studied how information of the degenerate sub-sector imprints on the behavior of the the sub-leading connected form factor. There, we showed that the onset time of the plateau and also the slope of the ramp are sensitive to the value of $\Gamma$ through the presence of the gap in the spectrum. In particular, in the double scaled matrix model examples given in Sections \ref{BesselModFFSection} and \ref{N=2Section}, we showed that the slope of the ramp depended on the the gap size and was given as $\frac{e^{-2\beta E_{\text{Gap}}}}{4\pi\beta}$. This result was recovered from a sine-kernel that emerges from a specific truncation of the non-perturbative kernel in the limit as $\hbar\to 0$. Using the sine-kernel truncation we also provided predictions on how the ramp would transition to the plateau.

In light of the results discussed above there are a number of interesting avenues of research to pursue. The first is from the perspective that the spectral form factor can be treated as a simple proxy to understand the thermalization behavior of a system which can be understood as random matrix models \cite{Cotler:2016fpe,Balasubramanian:2016ids,Collier:2021rsn,Saraswat:2021ong}, which we discussed in the introduction. In particular, given the oscillations we found in the spectral form factor which signaled the existence of a gap between the degenerate ground states and excited spectrum, it is natural to ask the question if similar features will also appear in computations of 2-point correlators of simple operators (similar to how in \cite{Cotler:2016fpe} certain correlators in the SYK model exhibited a ramp and plateau just like the spectral form factor did for those models). This is interesting to investigate especially in the context gravity and black holes. In particular, a 2-point correlator would represent how a perturbation thrown into a black hole with a spectral gap would be absorbed and thermalize. We would expect some kind of (potentially large) corrections to the thermalization behavior of a black hole in the ground state excited by a perturbation of characteristic energy of the size of the gap. Such an investigation would be similar in spirit to recent investigations of how Hawking radiation is affected by the spectral gap \cite{Lin:2025wof}. Perhaps one might attribute the oscillations in the form factor we found as a kind of partial reflection of an ingoing perturbation due to the lack of states in the gap resulting in a kind of ``echo'' \cite{Saraswat:2021ong}.  

Another interesting avenue of investigation is to more carefully study the non-perturbative terms we threw out to obtain the sine-kernel truncation to see if they can be organized in some general systematic framework and how it might connected to recent discussions of of corrections to the leading order $\tau$-scaling limit of the form factor \cite{Okuyama:2023pio}. Presumably, a more clear understanding of this may help us understand the early time behavior of the form factor which neither the leading truncation we used nor the $\tau$-scaling computation of the form factor can exactly reproduce. It appears that the $\tau$-scaling limit is designed to address only the late and intermediate transition between ramp and plateau, however our approach although not yet as robust, has the potential to study all time scales as $\hbar \to 0$ and beyond making it an interesting avenue to pursue in future work.

Finally, in recent work \cite{Johnson:2026plw} it was argued that in general, double scaled random matrix models which exhibit a $\Gamma$ number of degenerate ground (BPS) states with a macroscopic gap can be effectively regarded as being described by a $\Gamma\times \Gamma$ random matrix at sufficiently low energies. This is quite an interesting result and it would be interesting to explore if this could also be connected to the finding that the disconnected and connected part of the form factor swaps dominance when $\braket{Z_R(2\beta)}\sim \Gamma^2$ which also happens when the temperature is sufficiently low and we are focusing on the low energy sector of the theory. Perhaps the swapping of the dominance between the connected and disconnected part of the form factor might say something interesting about the transition from the full $(N+\Gamma)\times (N+\Gamma)$ matrix model description to the the $\Gamma\times\Gamma$ description in \cite{Johnson:2026plw}?

\acknowledgments
I would like to thank Clifford Johnson for his guidance during the course of this investigation as well as his valuable comments on the manuscript. I also thank Maciej Kolanowski, Mykhaylo Usatyuk, and Donald Marolf for helpful conversations and discussions. This work was supported by the University of California Santa Barbara and the US Department of Energy under the grant DE-SC 0011687.

\appendix

\section{BPS Microstates do not Contribute to $N$-Boundary Wormholes}
\label{NboudaryWHProof}
Here we provide a straightforward generalization of the discussion in Section \ref{SFFGenericAnalysisRandom+NonRandSec} which demonstrates that completely connected wormhole configurations connecting $N>2$ boundaries is insensitive to the presence of BPS microstates. The key to proving this is to note that $N$-boundary wormholes (fully connected) are in fact joint cumulants of the partition function \cite{Maldacena:2004rf,Saad:2019lba,Marolf:2020xie}. Therefore, to prove that the BPS sector does not contribute to these $N$-boundary wormholes it suffices to show the joint cumulant of the total partition function (which includes the BPS states) reduces to a joint cumulant of the partition function on the non-BPS sector alone. 

We can prove this directly by appealing to the generating function of the joint cumulant. Let $Z_i=\Gamma+Z_R(\beta_i)$, where $Z_R(\beta_i)$ is a random variable associated to the partition function of the non-BPS sector and $\Gamma$ is the partition function of the BPS sector. The generating function of the joint cumulant, denoted by $K_{\textbf{Z}}(\textbf{t})$, is given as:
\begin{equation}
    K_{\textbf{Z}}(\textbf{t})=\ln\left[\braket{e^{\sum_{i=1}^N t_i Z_i}}\right]\ .
\end{equation}
Now we can explicitly substitute $Z_i=\Gamma+Z_R(\beta_i)$ and use the key fact that $\Gamma$ is not a random variable to write:
\begin{equation}
     K_{\textbf{Z}}(\textbf{t})=\Gamma\sum_{i=1}^N t_i+\ln\left[\braket{e^{\sum_{i=1}^N t_i Z_R(\beta_i)}}\right]\ ,
\end{equation}
The $N$-th joint cumulant denoted as, $\kappa_N$, is given by differentiating the generating function with respect to $t_i$'s and then setting the $t_i$'s to zero as shown below\footnote{Note that in the notation we used in the main text of the paper $\braket{Z_1Z_2\cdot\cdot\cdot Z_N}_{con.}=\kappa_N(Z_1,Z_2,..,Z_N)$. In other words, the connected part of the form factor is exactly $\kappa_N(Z_1,Z_2)$ with $\beta_1=\beta+it$ and $\beta_2=\beta-it$.}:
\begin{equation}
    \kappa_N(Z_1,Z_2,..,Z_N)=\frac{\partial^N}{\partial t_1\partial t_2\cdot\cdot\cdot \partial t_N}  K_{\textbf{Z}}(\textbf{t})\vert_{\textbf{t}=0}\ .
\end{equation}
Applying this to our specific generating function we obtain:
\begin{equation}
    \begin{split}
        &\kappa_N(Z_1,Z_2,..,Z_N)= \frac{\partial^N}{\partial t_1\partial t_2\cdot\cdot\cdot \partial t_N}\left[\Gamma\sum_{i=1}^N t_i+\ln\left[\braket{e^{\sum_{i=1}^N t_i Z_R(\beta_i)}}\right]\right]\Bigg\vert_{\textbf{t}=0}\\
        &=\Gamma \delta_{1,N}+\frac{\partial^N}{\partial t_1\partial t_2\cdot\cdot\cdot \partial t_N} \left(\ln\left[\braket{e^{\sum_{i=1}^N t_i Z_R(\beta_i)}}\right]\right)\bigg\vert_{\textbf{t}=0}\ .\\
    \end{split}
\end{equation}
What we clearly see is that the first term involving $\Gamma$ is vanishes when $N\geq 2$ and the remaining term is exactly the joint cumulant of the partition function associated to the non-BPS sector which proves our statement that BPS states do not contribute to completely connected $N$-boundary wormholes in the gravitational path integral\footnote{In fact, these results can be viewed as a recasting of results discussed in the baby universe and $\alpha$-sectors context \cite{Gidding88,Marolf:2020xie} where the BPS states are a sector of the theory that do not vary among $\alpha$ sectors and thereby do not contribute to emission and absorption processes of baby universes.}. 

In many respects, the derivation provided here of the BPS sector not contributing to higher cumulants (i.e. wormhole geometries in the path integral) of the full partition function provides a amusing complementary view to the discussion of supersymmetric indices factorizing \cite{Iliesiu:2021are}. In the work \cite{Iliesiu:2021are}, the authors did a path integral computation with boundary conditions corresponding to computations of indices rather than the full partition function. There they showed that any connected wormhole contributions were exactly zero for the index. Our result is consistent with this computation although our starting point and approach was different. Instead of the index, we work with the full partition function which includes both BPS and non-BPS sectors and showed that the fully connected correlators of the partition function (i.e. cumulants) get no contribution from the BPS sector which effectively implies that the index (a computation of partition function restricted to BPS sector) must not contain any wormhole contributions. Furthermore, any non-factorization that occurs in the full partition function is solely because of the random non-BPS sector\footnote{The discussion of this appendix is only saying that BPS states do not contribute to wormholes in the context of computations of the ordinary partition function where the energy of BPS states do not fluctuate under different realizations of the ensemble. One could imagine constructing a different version of the partition function which is $e^{-\beta Q}$ where $Q$ is not necessarily the Hamiltonian operator. If $Q$'s action on the BPS states ``sees'' them as fluctuating with the ensemble then BPS states in this context will contribute to wormholes.}.

\section{Review of Aspects of Random Matrix Kernel}
\label{ReviewRMTKernel}
Consider the JPDF of the random eigenvalues of the complex Wishart ensemble given by \cite{hsu1939distribution,forrester1993laguerre,livan2018introduction,Johnson:2024tgg}:
\begin{equation}
    \rho(\lambda_1,..,\lambda_{N})=\frac{1}{\mathcal{Z}_N}\prod_{j<k}|\lambda_j-\lambda_k|^2\prod_{i=1}^N\lambda_i^{\Gamma}e^{-\sum_{p=1}^N\lambda_p}\ ,
\end{equation}
where $\mathcal{Z}_N$ is the partition function which acts as normalization constant to ensure that $\int \rho(\lambda_1,..,\lambda_N)d\lambda_1\cdot\cdot\cdot d\lambda_N=1$. What we can see from the JPDF of the random eigenvalues is that information of the $\Gamma$ degenerate sector appears as $\lambda_i^\Gamma$. The next thing to note is that the combination of terms $\prod_{j<k}|\lambda_i-\lambda_k|^2=\left|\text{det}\left[\lambda_j^{i-1}\right]\right|^2=|\Delta_N(\vec{\lambda})|^2$ is really a Vandermonde determinant. Then we may express the JPDF as:
\begin{equation}
    \rho(\lambda_1,..,\lambda_{N})=\frac{1}{\mathcal{Z}_N}|\Delta_N(\vec{\lambda})|^2\prod_{i=1}^N\lambda_i^{\Gamma}e^{-\sum_{p=1}^N\lambda_p}\ .
\end{equation}
Finally we can rewrite the remaining terms in the form $\prod_{i=1}^N e^{-V(\lambda_i)}$ with:
\begin{equation}
    V(\lambda_i)=\lambda_i-\Gamma\ln(\lambda_i)\ .
\end{equation}
This leaves us with an expression of the form:
\begin{equation}
\begin{split}
&\rho(\lambda_1,..,\lambda_{N})=\frac{1}{\mathcal{Z}_N}\prod_{i=1}^N e^{-V(\lambda_i)}|\Delta_N(\vec{\lambda})|^2\ .\\
    \end{split}
\end{equation}
The next step is to note that the Vandermonde determinant is invariant up to a rescaling under redefinition of entries $\lambda_j^{i-1}$ with a polynomial $P_{i-1}(\lambda_{j})=a_{i-1}\lambda_j^{i-1}+\cdot\cdot\cdot$. For such a matrix one would get:
\begin{equation}
|\Delta_N(\vec{\lambda})|^2=\left[\frac{\text{det}\left[P_{i-1}(\lambda_j)\right]}{a_0a_1\cdot\cdot\cdot a_{N-1}}\right]^2=\frac{1}{\prod_{i=0}^{N-1}a_i^2}\det\left[\sum_{p=1}^NP_{p-1}(\lambda_i)P_{p-1}(\lambda_j)\right]\ .
\end{equation}
After pulling the $\prod_{i=1}^Ne^{-V(\lambda_i)}$ inside the determinant gives the following:
\begin{equation}
\begin{split}
&\rho(\lambda_1,..,\lambda_{N})=\frac{\text{det}\left[K_N(\lambda_i,\lambda_j)\right]}{\mathcal{Z}_N\prod_{i=1}^{N-1}a_i^2}\\
&K_N(\lambda_i,\lambda_j)=e^{-\frac{1}{2}V(\lambda_i)-\frac{1}{2}V(\lambda_j)}\sum_{p=0}^{N-1}P_{p}(\lambda_i)P_{p}(\lambda_j)\ .\\
\end{split}
\end{equation}
The object $K_N(\lambda_i,\lambda_j)$ is the sought after object called the kernel. It is particularly useful when one defines $P_k(\lambda_i)$ as orthonormal polynomials which satisfy the identity:
\begin{equation}
    \int e^{-V(x)}P_k(x)P_\ell(x)dx=\delta_{k\ell}\ .
\end{equation}
Now that we have motivated and introduced the kernel we can see that the reproducing property given in Eq. (\ref{ReproducingPropOfKernel}) enables the explicit evaluation of integrals when computing various marginal probability distributions of $n<N$ eigenvalues. In particular, the spectral density is obtained by integrating out $N-1$ eigenvalues:
\begin{equation}
    \tilde{\rho}(\lambda)=\int \rho(\lambda,\lambda_2,...,\lambda_N)d\lambda_2,d\lambda_3\cdot\cdot\cdot d\lambda_N=\frac{1}{N}K_N(\lambda,\lambda)\ .
\end{equation}
It demonstrates that the normalized spectral density is actually the diagonal of the kernel which we can express in terms of a sum of orthogonal polynomials.

\section{Derivation of Expression for $C_J(N,\Gamma)$}
\label{DeriveFormulaCJAppendix}
In this appendix we derive the formula in Eq. (\ref{FormulaForCJ}) for $C_J(N,\Gamma)$ which appears in the definition of the partition function for the Wishart ensemble. To begin, we recall that the definition of the coefficient given in Eq. (\ref{DefiitionofCJ}). We will make use of the following summation representation of Laguerre polynomials:
\begin{equation}
    L_N^{(\Gamma)}(x)=\sum_{k=0}^N\frac{(-1)^k}{k!}\binom{N+\Gamma}{N-k}x^k\ .
\end{equation}
It follows that:
\begin{equation}
    \begin{split}
        &L_{N-1}^{(\Gamma)}(x)L_{N-1}^{(\Gamma+1)}(x)=\sum_{p=0}^{N-1}\sum_{k=0}^{N-1}\frac{(-1)^{k+p}}{k!p!}\binom{N+\Gamma-1}{N-p-1}\binom{N+\Gamma}{N-k-1}x^{p+k}\ .\\
    \end{split}
\end{equation}
Defining $J=p+k$ we can rewrite the double sum as follows:
\begin{equation}
    L_{N-1}^{(\Gamma)}(x)L_{N-1}^{(\Gamma+1)}(x)=\sum_{J=0}^{2N-2}\sum_{k=0}^J\frac{(-1)^J}{k!(J-k)!}\binom{N+\Gamma-1}{N+k-J-1}\binom{N+\Gamma}{N-k-1}x^J\ .
\end{equation}
Using analogous method we can also write:
\begin{equation}
    L_N^{(\Gamma)}(x)L_{N-2}^{(\Gamma+1)}(x)=\sum_{J=0}^{2N-2}\sum_{k=0}^J\frac{(-1)^J}{k!(J-k)!}\binom{N+\Gamma}{N-k}\binom{N+\Gamma-1}{N+k-J-2}x^J\ .
\end{equation}
Combining these results gives:
\begin{equation}
\begin{split}
    &L_{N-1}^{(\Gamma)}(x)L_{N-1}^{(\Gamma+1)}(x)-L_N^{(\Gamma)}(x)L_{N-2}^{(\Gamma+1)}(x)\\
    &=\sum_{J=0}^{2N-2}\sum_{k=0}^J\frac{(-1)^J}{k!(J-k)!}\left[\binom{N+\Gamma}{N-k-1}\binom{N+\Gamma-1}{N+k-J-1}-\binom{N+\Gamma}{N-k}\binom{N+\Gamma-1}{N+k-J-2}\right]x^{J}\ .\\
\end{split}
\end{equation}
We have written the expression in Eq. (\ref{DefiitionofCJ}) in the form of the right-hand side of the equation we can simply read off:
\begin{equation}
    C_J(N,\Gamma)= \sum_{k=0}^J\frac{(-1)^J}{k!(J-k)!}\left[\binom{N+\Gamma}{N-k-1}\binom{N+\Gamma-1}{N+k-J-1}-\binom{N+\Gamma}{N-k}\binom{N+\Gamma-1}{N+k-J-2}\right]\ .
\end{equation}
This sum can be done exactly and results in the expression in Eq. (\ref{FormulaForCJ}) in terms of Hypergeometric functions.

\bibliography{Ref.bib}
\bibliographystyle{JHEP}

\end{document}